\documentclass[%
 reprint,
 superscriptaddress,
 amsmath,amssymb,
 aps,
]{revtex4-2}

\usepackage{natbib}
\usepackage{tikz,pgfplots}
\usepackage{graphicx}
\usepackage[caption=false]{subfig}
\usepackage{multirow}
\usepackage{float}
\usepackage{dcolumn}
\usepackage{bm}

\begin{document}

\preprint{APS/123-QED}

\title{Imaging electron angular distributions to assess a full-power petawatt-class laser focus}

\author{Smrithan Ravichandran}
\affiliation{%
 Institute for Physical Science and Technology, University of Maryland, College Park, Maryland 20742, USA
}%
\affiliation{Joint Quantum Institute, University of Maryland, College Park, Maryland 20742, USA}

\author{Marine Huault}
\affiliation{Centro de L\'aseres Pulsados, 37185 Villamayor, Salamanca, Spain}
\affiliation{Departamento de F\'isica Fundamental, Universidad de Salamanca, 37008 Salamanca, Spain}

\author{Roberto Lera}
\affiliation{Centro de L\'aseres Pulsados, 37185 Villamayor, Salamanca, Spain}

\author{Calvin Z. He}
\affiliation{%
 Institute for Physical Science and Technology, University of Maryland, College Park, Maryland 20742, USA
}%
\affiliation{Joint Quantum Institute, University of Maryland, College Park, Maryland 20742, USA}

\author{Andrew Longman}%
\affiliation{Lawrence Livermore National Laboratory, Livermore, California 94550, USA}

\author{Robert Fedosejevs}
\affiliation{Electrical and Computer Engineering, University of Alberta, Edmonton, Alberta T6G 2V4, Canada}

\author{Luis Roso}
\affiliation{Departamento de F\'isica Aplicada, Universidad de Salamanca, 37008 Salamanca, Spain}

\author{Wendell T. Hill III}
\email{wth@umd.edu}
\affiliation{%
 Institute for Physical Science and Technology, University of Maryland, College Park, Maryland 20742, USA
}%
\affiliation{Joint Quantum Institute, University of Maryland, College Park, Maryland 20742, USA}
\affiliation{Department of Physics, University of Maryland, College Park, Maryland 20742, USA}

 




\begin{abstract}
We present a novel technique to assess the focal volume of petawatt-class lasers at full power. Our approach exploits quantitative measurement of the angular distribution of electrons born in the focus via ionization of rarefied gas, which are accelerated forward and ejected ponderomotively by the field. We show that a bivariate ($\theta, \phi$) angular distribution, which was obtained with image plates, not only enables the peak intensity to be extracted, but also reflects nonideality of the focal-spot intensity distribution. In our prototype demonstration at intensities of a few $\times 10^{19}$ to a few $\times 10^{20}$ $\mathrm{W/cm^2}$, an f/10 optic produced a focal spot in the paraxial regime. This allows a plane-wave parameterization of the peak intensity given by $\tan{\theta_c} = 2/a_0$ ($a_0$ being the normalized vector potential and $\theta_c$ the minimum ejection angle) to be compared with our measurements. Qualitative agreement was found using an $a_0$ inferred from the pulse energy, pulse duration and the focal spot distribution with a modified parameterization, $\tan{\theta_c} = 2\eta/a_0$ ($\eta = 2.02^{+0.26}_{-0.22}$). This highlights the need for (i) better understanding of intensity degradation due to focal-spot distortions and (ii) more robust modeling of the ejection dynamics. Using single-shot detection of electrons, we showed that while there is significant shot-to-shot variation in the number of electrons ejected at a given angular position, the average distribution scales with the pulse energy in a way that is consistent with that seen with the image plates. Finally, we note that the asymptotic behavior as $\theta \to 0^{\circ}$ limits the usability of angular measurement. For 800 nm, this limit is at an intensity $\sim 10^{21}\ \mathrm{W/cm^2}$. 

\end{abstract}

\maketitle


\section{\label{sec:level1}Introduction}

The recent proliferation of petawatt-class laser facilities \cite{danson_petawatt_2015} is driven by a two-pronged desire -- creating extreme intensities in the laboratory \cite{yoon_realization_2021,yoon_ultra-high_2022} where new fundamental physics can be explored, and developing new laser-based technologies \cite{piazza_multi-petawatt_nodate}. Pair production \cite{bell_possibility_2008,krajewska_bethe-heitler_2013,zhu_dense_2016,salgado_towards_2021} and strong-field photon-photon scattering \cite{karplus_non-linear_1950,costantini_nonlinear_1971,gies_photon-photon_2018,roso_towards_2022} are examples of the former. Secondary sources, particle accelerators and high-energy photon beams, are examples of the latter. While efforts are underway to upgrade or build bigger and more powerful lasers \cite{noauthor_zeus_nodate,noauthor_clf_nodate,bromage_technology_2019,noauthor_apollon_nodate,radier_10_2022,peng_yujie_overview_2021}, instrumentation to characterize the focal spot, necessary to guide and improve designs, and to employ as experimental diagnostic tools, have not kept pace. Intensity estimates today, as it has been for some years, still largely rely on indirect approaches that either do not sample the full beam at full power or are not performed in real time. As such, these estimates tend to depend on extrapolations and assumed behavior of laser parameters that fail to account for real-time beam conditions and fluctuations, intensity-dependent degradation due to spatiotemporal coupling \cite{akturk_spatio-temporal_2010}, beam aberrations and other nonlinear effects in the focus. Pulse-front tilts as small as 0.2 fs/mm due to slight imperfections in compressor gratings, for example, have been shown to reduce the intensity as much as an order of magnitude \cite{ouatu_ionization_2022}. Clearly, new tools are needed.

Several approaches have been suggested to characterize the peak intensity at full power, exploiting concomitant processes with high intensities ($> 10^{18}\ \mathrm{W/cm^2}$) -- radiation associated with relativistic Thomson scattering \cite{harvey_situ_2018,gao_laser_2006,har-shemesh_peak_2012,he_towards_2019}, appearance intensity of ionization stages of tenuous gases \cite{link_development_2006,ciappina_progress_2019} and ponderomotive ejection of electrons \cite{galkin_electrodynamics_2010,kalashnikov_diagnostics_2015,ivanov_accelerated_2018,mackenroth_ultra-intense_2019}. For widespread use as a diagnostic, it is important that the method be straightforward to implement, minimally intrusive to the principal scientific study, sensitive to beam conditions and distortions, and capable of single-shot deployment. Thomson scattering in the $10^{18}$ to $10^{19}\ \mathrm{W/cm^2}$ intensity range produces spectrally-convenient Doppler-shifted, $2^{\mathrm{nd}}$ harmonic radiation that is detectable with a gated spectrometer \cite{he_towards_2019}. The spectrometer requirement, however, makes single-shot deployment challenging. At the same time, intensities $> 10^{19} \ \mathrm{W/cm^2}$ cause harmonic orders to overlap making them difficult to distinguish. While monitoring the fundamental is possible, mid-infrared detection will be required. Ionization is single-shot capable but provides limited information on the intensity distribution in the focus. Moreover, ionization thresholds tend to be clumped together for certain intensities, with large gaps between thresholds at other intensities. This manuscript introduces a straightforward experimental technique upon which an in situ intensity-measurement tool might possibly be based. The approach exploits the angular distribution of electrons ejected from the focus. The electrons not only allow the intensity to be monitored, they reveal asymmetries in the intensity distribution within the focal volume. 

Figure \ref{subfig:IP_schematic} shows the experimental schematic of nascent electrons being ejected into a bivariate ($\theta$ and $\phi$) angular distribution (BiAD) that we measured using image plates. In distinction to previous studies, to the best of our knowledge, this is the first experiment to capture two-dimensional images (all $\phi$) of electrons, which reveals new information about the laser focus. The BiAD measurements were taken under paraxial conditions that allowed us to test the validity of the plane-wave parameterization discussed below. Our measurements were done at pressures between $10^{-5}$ to $10^{-4}$ mbar. In this range, the image plates required $\approx 100$ laser shots. We also made single-shot measurements of the electrons with scintillation detectors. 

This manuscript is organized as follows. We provide some background and outline the theory in Sec.~\ref{sec:theory_section}. In Sec.~\ref{sec:exp_methods_section}, we describe the experiments. We present the results from our analysis of the BiADs in Sec.~\ref{sec:discussion_section}. We also compare the measured $\tan{\theta}$ vs.~$a_0$ scaling with the theoretical predictions and discuss the limitations of this application for higher intensities. Finally, we present a roadmap for deployment as a tomographical tool and how it might be extended well beyond the current intensity record of $10^{23}\ \mathrm{W/cm^2}$ \cite{yoon_realization_2021}.

\section{Theory} \label{sec:theory_section}
It is well known that the ponderomotive force causes free electrons to experience an outward force from the focus of laser beams \cite{hartemann_nonlinear_1995,quesnel_theory_1998}. When the normalized vector potential, $a_0 = eE\lambda_0/2\pi m_e c^2 \approx 0.855 \lambda_0$($\mu$m)$\sqrt{I (\mathrm{W/cm^2})/10^{18}}$, exceeds $1$, the laser accelerates electrons to relativistic energies within a single cycle. Here, $e, \ E, \ \lambda_0, \ m_e, \ c, \ I$ are the elementary charge, electric field magnitude, laser wavelength, electron mass, speed of light in vacuum and peak laser intensity respectively. Consequently, the magnetic and electric forces become comparable, resulting in the electrons gaining momentum along the direction of laser propagation. Due to the existence of longitudinal field components, as necessitated by Maxwell's equations, the electron ejection is not restricted to a plane \cite{quesnel_theory_1998}. Rather, electrons with a Lorentz factor $\gamma$, are ejected into a cone about the wave vector, $\vec{k}$, with apex angle $\theta(\gamma)$ that obeys \cite{moore_observation_1995,hartemann_nonlinear_1995}
\begin{equation}\label{eqn:tan_theta}
    \tan\theta = p_\perp/p_\parallel = \sqrt{2/(\gamma-1)},
\end{equation}

\noindent where $p_{\perp}$ and $p_{\parallel}$ are the transverse and longitudinal components of the electron's momentum relative to $\vec{k}$ respectively. As electrons interact with higher laser intensities, $\gamma$ increases, which decreases $p_{\perp}/p_{\parallel}$ and $\theta$ according to Eq.~(\ref{eqn:tan_theta}). Thus, one might expect the existence of a characteristic ejection cutoff angle, $\theta_{c}$, for the BiAD of ejected electrons that reflects the peak intensity, or $a_0$, experienced by the electrons. The exact dependence of $\theta_c$ on $a_0$, however, requires knowledge of how the kinetic energy of the most energetic electrons with $\gamma = \gamma_p$ is related to $a_0$. While studies have considered this relationship in various contexts, how the ponderomotive force contributes to $\gamma_p$ is not known analytically for a general case due to theoretical complications in treating the relativistic dynamics of the accelerated electrons. A plane-wave analysis by Hartemann et al. \cite{hartemann_nonlinear_1995}, predicts

\begin{equation}
     \gamma_p = 1 + a^{2}_{0}/2 \label{eqn:gamma_hartemann}
\end{equation}

\begin{figure}
  \centering
  \subfloat[\label{subfig:IP_schematic}]{\includegraphics[width=1\linewidth]{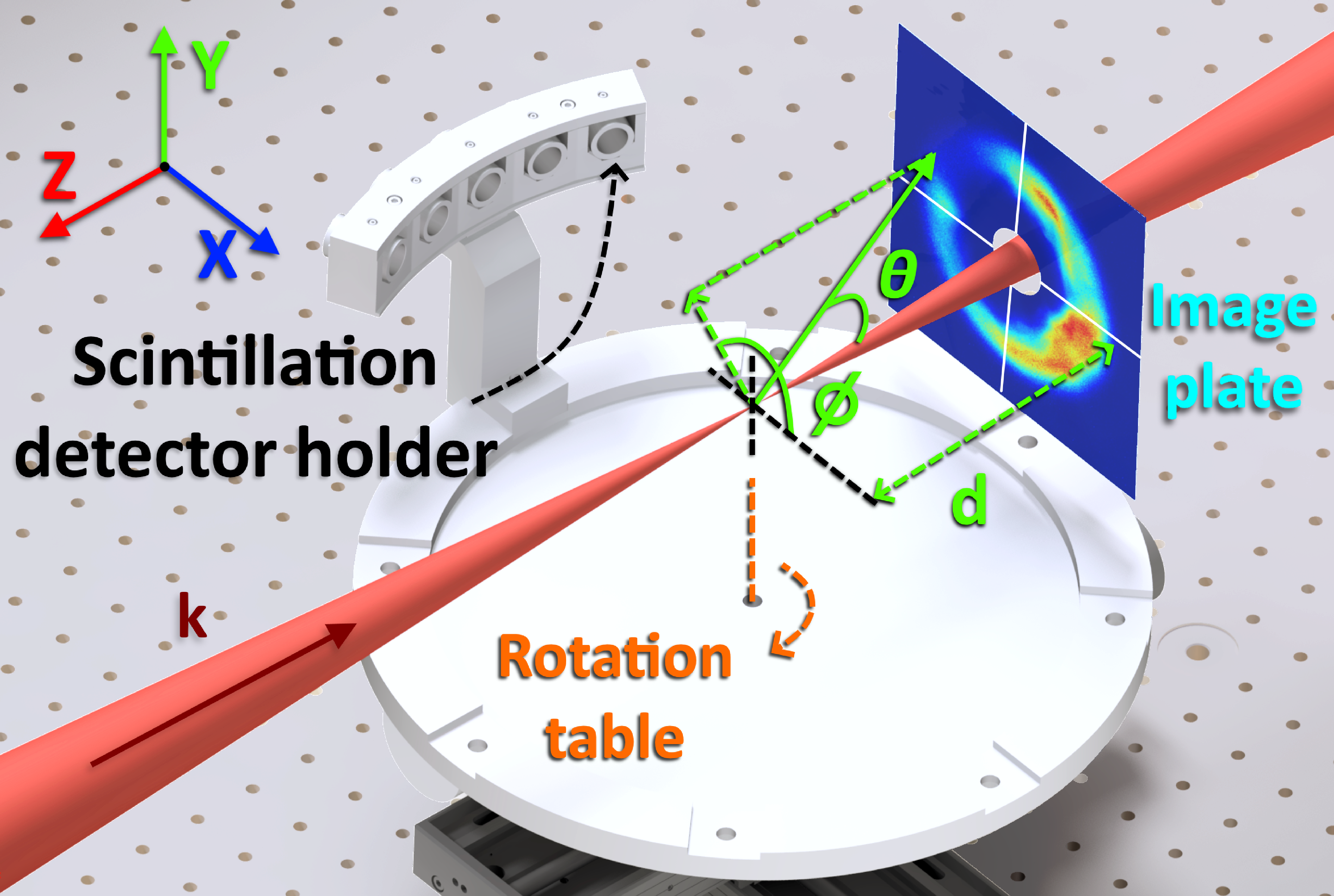}}\\

  \subfloat[\label{subfig:focal_spot}]{\includegraphics[height=3.7cm,trim={0.2cm 0 3.4cm 0.5cm},clip]{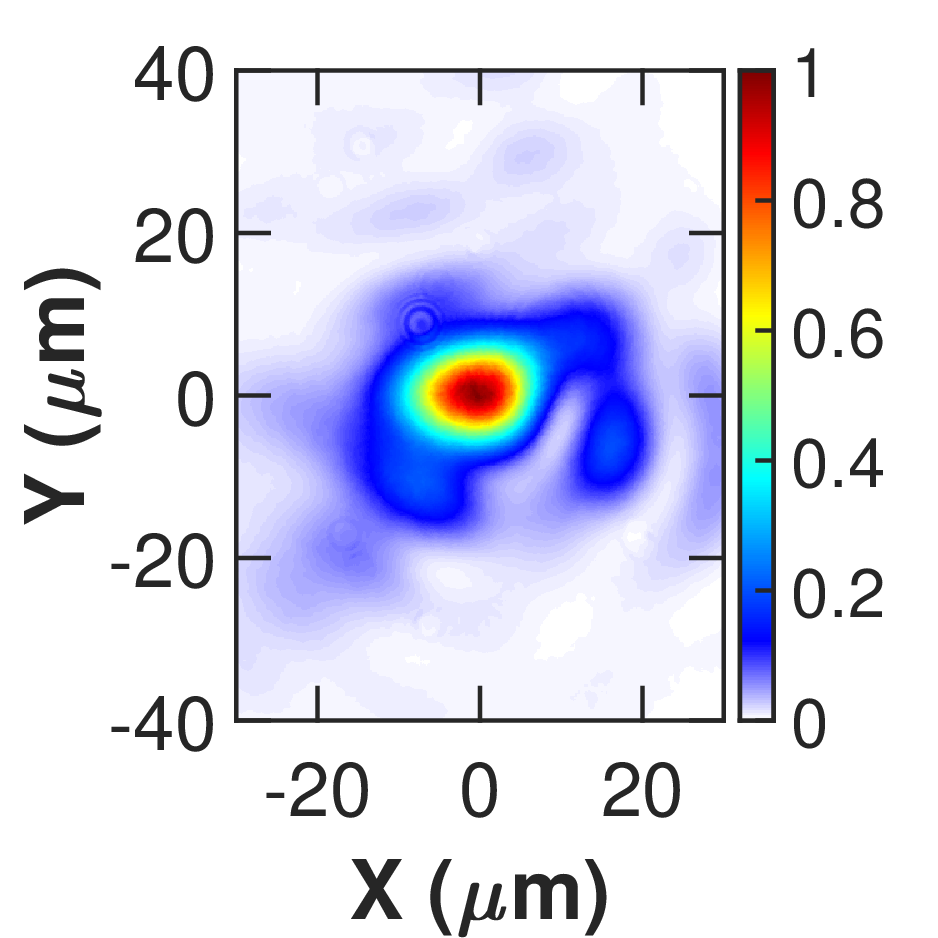}}\hspace{0mm}
  \subfloat[\label{subfig:focal_spot_2}]{\includegraphics[height=3.7cm,trim={4cm 0 3.4cm 0.5cm},clip]{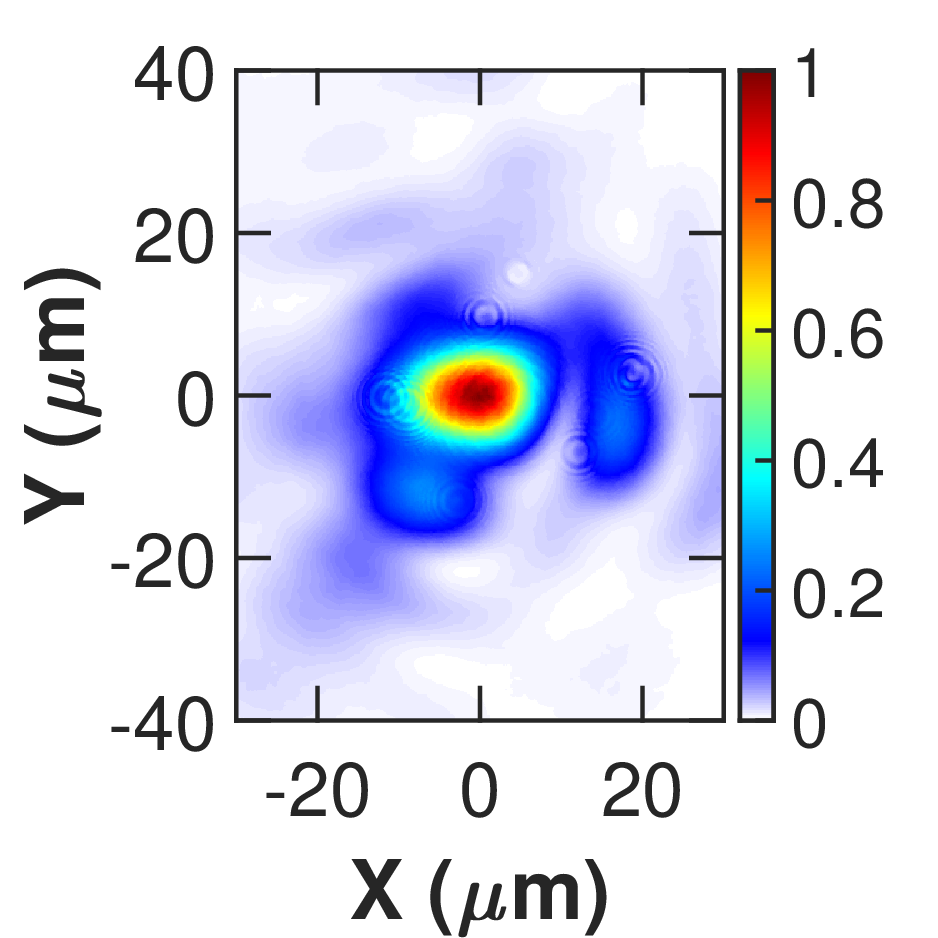}}\hspace{0mm}
  \subfloat[\label{subfig:focal_spot_3}]{\includegraphics[height=3.7cm,trim={4cm 0 3.4cm 0.5cm},clip]{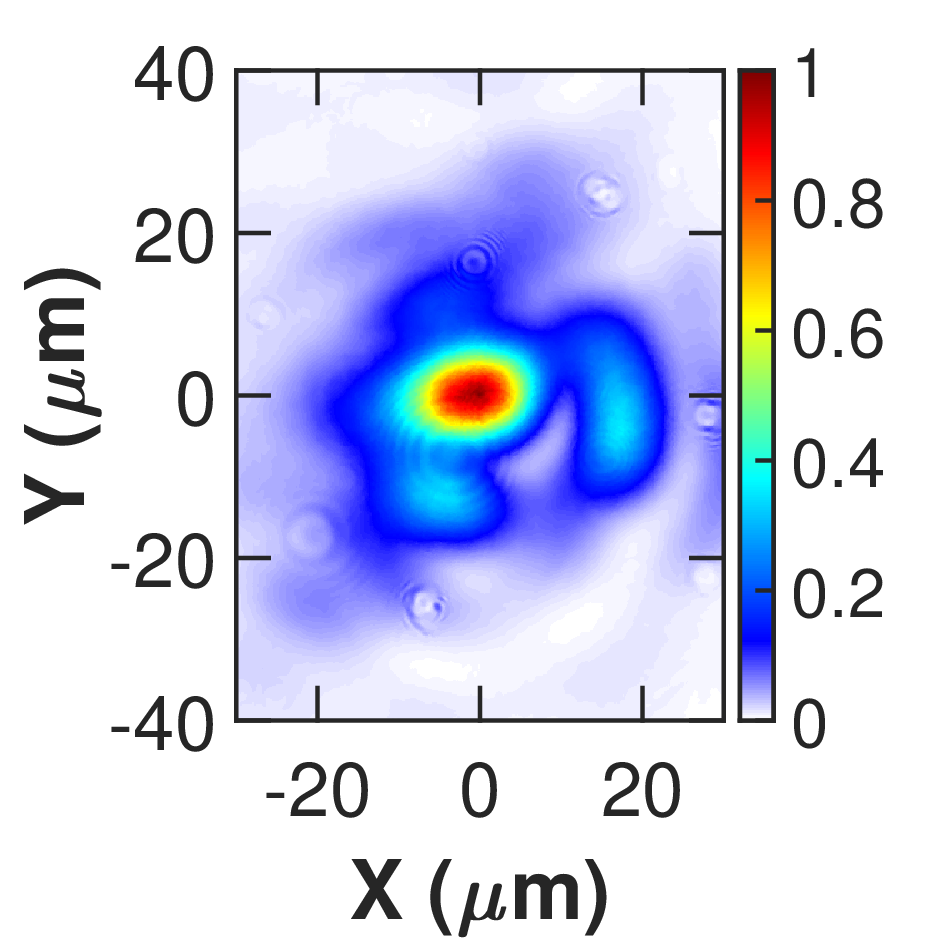}} \hspace{1mm}
  \subfloat{\includegraphics[height=3.7cm,trim={12.5cm 0 0 0.5cm},clip]{Focal_spot_imaged_at_low_power_3_low.eps}}\rotatebox{90}{\hspace{1.2cm}\textbf{\fontfamily{phv}\selectfont{Scaled signal}}} \\ 
  \caption{Experimental setup (not to scale) for image plate and scintillation detectors (a). Image plates are mounted a fixed distance, $d\approx 40$ mm, after the focus with active area facing the focal spot. Scintillation detectors (not shown) are mounted in the holder, attached to a rotation table, rotating about the focus in the $\vec{E}$-$\vec{k}$ (i.e., $\hat{x}$-$\hat{z}$) plane a fixed distance ($\approx 138$ mm) away. Typical VEGA-3 focal spot images obtained at low power are shown in (b)--(d).}
  \label{fig:Exp_setup}
\end{figure}

\noindent for the interaction of free electrons with a linearly polarized field \cite{kalashnikov_diagnostics_2015,ivanov_accelerated_2018,salamin_ponderomotive_1997}. We note that for a focused laser pulse travelling along $-\hat{z}$ and polarized with $E_x$ along $\hat{x}$ as shown in Fig.~\ref{subfig:IP_schematic}, the longitudinal component of the electric field, $E_z$, has a first-order contribution $\propto$ $\partial E_x / \partial x$ \cite{erikson_polarization_1994,pang_subluminous_2002,longman_modeling_2022}. For a TEM$_{00}$ Gaussian laser mode, it can be shown that the largest contribution scales as $E_z(r=w_0/\sqrt{2})/E_x(r=0) \approx 1/(kw_0$) \cite{erikson_polarization_1994}, where $r$ is the radial distance in the transverse plane and $w_0$ is the beam waist of the Gaussian laser beam. Therefore, as the f/\# (f-number) of the focusing optic increases, $E_z/E_x$ decreases. In our experiment, we used an off-axis f/10 parabolic mirror where $1/(kw_0) \approx 0.018$, which is in the paraxial limit.

We note there are other relationships between $\gamma_p$ and $a_0$; $\gamma_p = \sqrt{1 + a^{2}_{0}/2}$ \cite{wilks_absorption_1992} is often considered in the context of energy absorption from a high-intensity laser pulse by an electron in a plasma \cite{cai_short-pulse_2006,cui_laser_2013}. It has also been considered for free electrons in a laser field \cite{ciappina_progress_2019}. This scaling and that of Eq.~(\ref{eqn:gamma_hartemann}) differ significantly in the value of $a_0$ for a given $\gamma_p$. For example, for $\gamma_p = 2$, $a_0$ according to Eq.~(\ref{eqn:gamma_hartemann}) is $60\%$ of that in $\gamma_p = \sqrt{1 + a^{2}_{0}/2}$. Thus, it is important to distinguish between conditions under which either is applicable, if at all, and test them to gain a better understanding of the role of the ponderomotive force in this relativistic intensity regime. While a direct quantitative measurement of $a_0$ may be complicated, we can test the plane-wave prediction of the dependence of $\theta_c$ as a function of $a_0$ by combining Eqs.~(\ref{eqn:tan_theta}) and (\ref{eqn:gamma_hartemann}) to give

\begin{equation}
     \tan{\theta_c} = 2/a_0 \label{eqn:tan_theta_a_0}.
\end{equation}

\noindent To that end, we point out that $a_0$ depends on the pulse energy and pulse duration. Thus, by varying these parameters, we can compare the $\theta_c$ measured for varying $a_0$ inferred from low-power measurements to Eq.~(\ref{eqn:tan_theta_a_0}). In the next section, we demonstrate the effectiveness of the image plate technique as a simple, yet powerful diagnostic to measure the ring-like BiAD of ejected electrons and to find $\theta_c$. 

\section{Experimental methods}\label{sec:exp_methods_section}

Our experiment was performed on the VEGA-3 petawatt laser \cite{noauthor_clpu_nodate} at the Centro de L\a'aseres Pulsados (CLPU) in Spain, with $\lambda_0 = 0.8$ $\mu$m. A cartoon of the setup is shown in Fig.~\ref{subfig:IP_schematic}. We generated free electrons via the ionization of low density gases ($\approx$ 10$^{-5}$ -- 10$^{-4}$ mbar) to create sufficiently large electron signals that could be detected by our instrumentation. As the electrons of interest are accelerated to relativistic energies by the laser, gaining a final energy that is far greater than their energy at birth, we consider them to be at rest initially. The gas pressures were chosen low enough to reduce space charge and collective plasma effects, allowing us to work in the single-particle regime. A conservative estimate of the Debye shielding length \cite{chen_introduction_1984,bellan_fundamentals_2008}, $\lambda_D$, for example shows that even completely ionizing all 14 electrons of the nitrogen molecule at 10$^{-4}$ mbar ($\approx 2.4 \times 10^{18}$ molecules/m$^3$), $\lambda_D$ $\approx 0.3$ mm for 50-keV electrons and $\approx 0.4$ mm for 100-keV electrons. This is much larger than the size of the focal spot, and thus allows us to ignore potential collective effects. 

\subsection{Image plates}\label{ssec:IP_section}

\def\arraystretch{1.6}
\begin{table}
    \begin{center}
        \caption{Summary of experimental conditions used to measure $\theta_c$ with nitrogen gas. The $U$ and $\tau$ (as defined in Sec.~\ref{ssec:a_0_estimation}) reported here is the average value obtained over the corresponding sequence of shots, after considering the systematic uncertainties in the measurement process, which dominate the uncertainty in measuring the average. The techniques used to measure $U$ and $\tau$ are detailed in Sec.~\ref{ssec:a_0_estimation}. The sources for the uncertainty in $\theta_c$ are the error in positioning the image plate and in estimating where the signal cuts off as detailed in the text.\\}
        \begin{tabular}{l l l l l l l}     
            \hline\hline
            \multirow{2}{2em}{\textbf{Fig.}} & \multirow{2}{3em}{\textbf{Plate Type}} & \multirow{2}{3.5em}{\textbf{No.~of Shots}} & \multirow{2}{3.75em}{\textbf{U (J) (\boldmath $\pm 10\%$)}} & \multirow{2}{3.75em}{\textbf{\boldmath $\tau$ (fs) ($\pm 20\%$)}} & \multirow{2}{4.5em}{\textbf{Pressure (mbar)}} & \multirow{2}{3em}{\textbf{\boldmath $\theta_c$ (deg)}}\\ \\
            \hline
            \ref{subfig:IP_High_Int} & SR & 100 & 23.0 & 35.4 & $4\times10^{-5}$ & 22 $\pm$ 1\\ 
            \ref{subfig:IP_Med_Int} & SR & 100 & 23.4 & 55.4 & $7\times10^{-5}$ & 30 $\pm$ 1\\ 
            \ref{subfig:IP_Low_Int} & MS$^{\dag}$ & 96 & 8.8 & 34.7 & $9\times10^{-5}$ & 39 $\pm$ 2\\ 
            \hline\hline
            \multicolumn{7}{p{26em}}{$^{\dag}$\footnotesize{MS plates were used for the lower intensity measurement as they are more sensitive than the SR by a factor of $\approx 3$.}}
        \end{tabular}\label{tab:IP_params}
        \end{center}        
\end{table}

\begin{figure}
     \centering
     \vspace{0.5cm}
     \begin{tabular}{cc}  
        \hspace{-2.2cm}\subfloat[\label{subfig:IP_High_Int}]{\includegraphics[width=0.46\columnwidth]{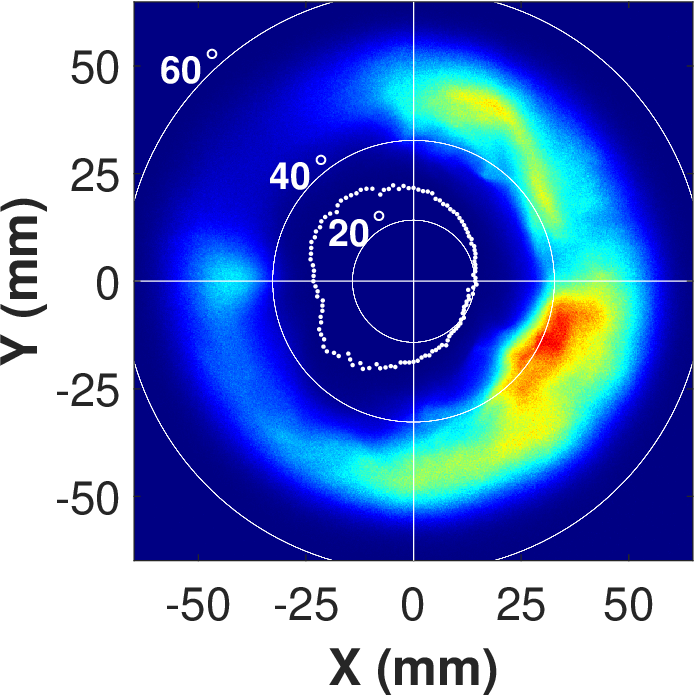}}&
        \hspace{-1.2cm}\subfloat[\label{subfig:IP_Med_Int}]{\includegraphics[width=0.46\columnwidth]{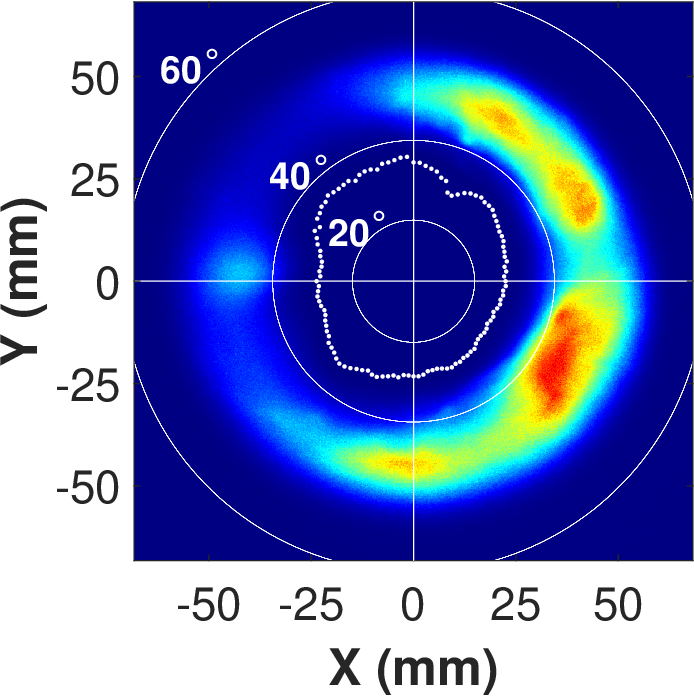}} \vspace{3mm}  \\  
        \hspace{1cm}\subfloat[\label{subfig:IP_Low_Int}]{\includegraphics[width=0.46\columnwidth]{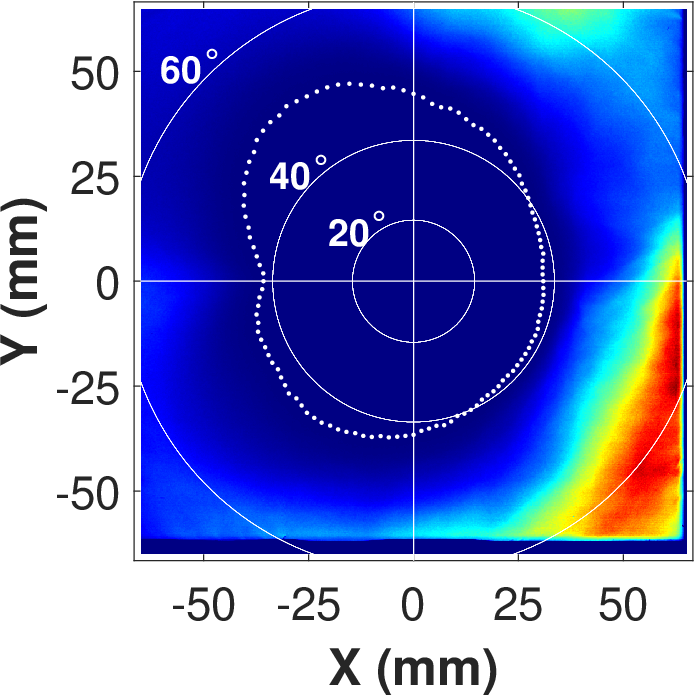}}& 
        \hspace{-1.5cm}\begin{tikzpicture}
             \begin{axis}[
             at={(-5,-0.5)},
             hide axis,
             scale only axis,
             height=4cm,
             width=1cm,
             colormap/jet,
             colorbar horizontal,
             colorbar right,
             point meta min=0,
             point meta max=1,
             colorbar style={
             width = 0.2cm,
             height = 3.25cm,
             ylabel = \textbf{\fontfamily{phv}\selectfont{Scaled signal}},
             ylabel style = {
                at={(12,0.5)}
             }
             }]
             \end{axis}
             \end{tikzpicture}
     \end{tabular}
     \caption{Scaled image plate data captured under conditions detailed in Table \ref{tab:IP_params}, where 1 on the colorbar corresponds to peak PSL values (see text) of 7.06, 6.82 and 31.88 respectively for (a)--(c). The laser is polarized along the $\vec{x}$-axis and $\vec{k}$ points into the page. White ``dots" indicate the single-electron PSL level vs $\phi$ (see text).}
     \label{fig:IP_scans}
\end{figure}

We used two types of commercially available image plates -- Fujifilm BAS-MS and BAS-SR -- to image the BiAD of ejected electrons under different experimental conditions specified in Table \ref{tab:IP_params}; images are shown in Fig.~\ref{fig:IP_scans}. Image plates store a fraction of the energy of traversing electrons in a photo stimulable phosphor layer that can be read with a phosphorimager device \cite{takahaahi_mechanism_nodate,takahashi_photostimulated_1985,miyahara_new_1986}. Plates with a hole (radius $\approx$ 10 mm) to allow the laser to pass through were mounted $\approx 40 \pm 2$ mm after the focus, perpendicular to $\vec{k}$, and facing the focal spot as shown in Fig.~\ref{subfig:IP_schematic} (not to scale). They were covered with a single layer of Al foil (thickness 12 $\mu$m) to block scattered laser light. In the case of Figs.~\ref{subfig:IP_High_Int} and \ref{subfig:IP_Med_Int}, we placed an additional Al shield (thickness 520 $\mu$m) directly in front of the plates to block the abundant low energy electrons to enhance the contrast for the most energetic electrons of interest. This increases the detection threshold to $\approx$ 0.5 MeV \cite{noauthor_stopping_nodate} and sets the large-angle falloff to $\approx 55^{\circ}$ according to Eq.~(\ref{eqn:tan_theta}), which is in good agreement with the observed large-angle falloff for the measurements as shown. After correcting for the signal fading in time \cite{boutoux_study_2015} before the plate is read, all measurements are shown in photostimulated luminescence (PSL) units \cite{williams_calibration_2014}, which is linearly proportional to the energy deposited on the plate. 

\begin{figure}
     \centering
     \includegraphics[width=1\linewidth]{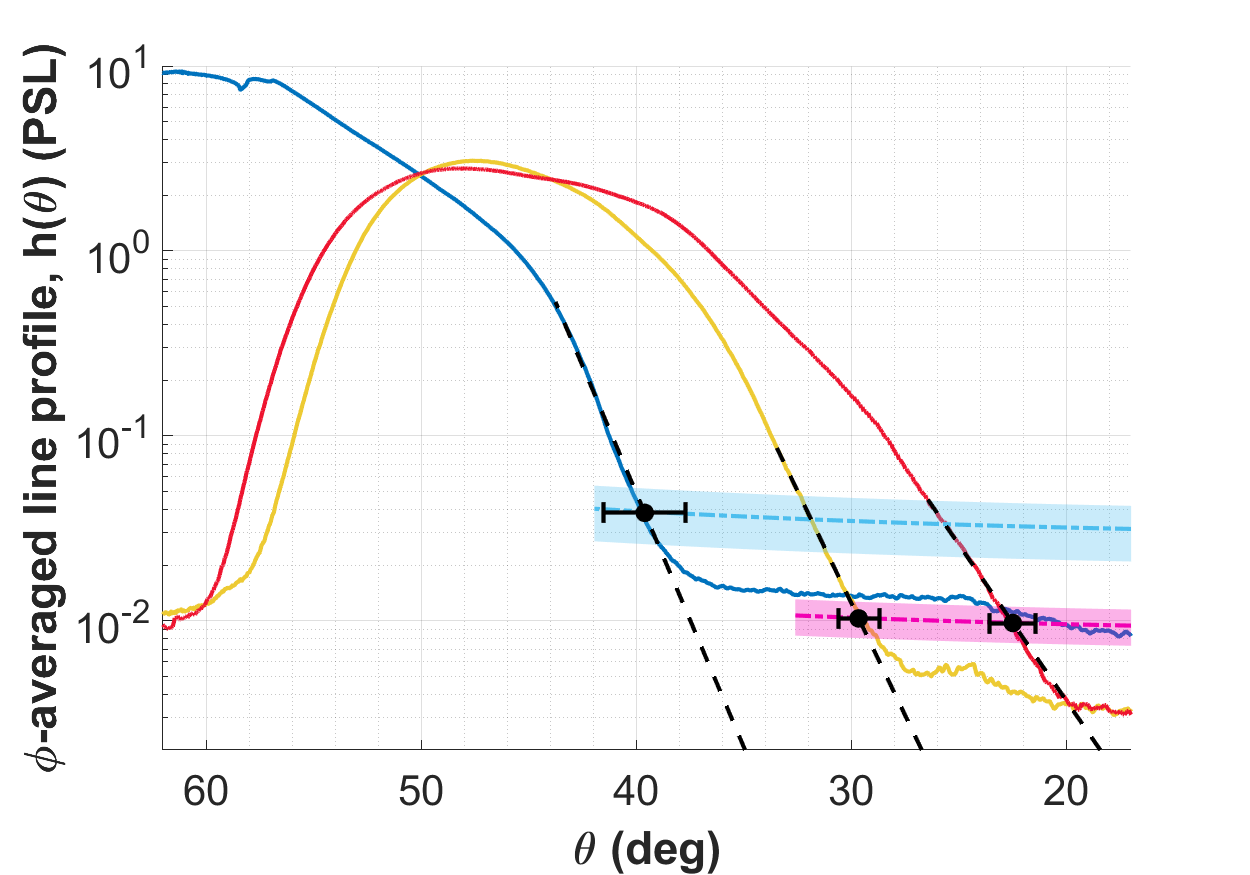}
     \caption{The $\phi$-averaged line profile, $h(\theta)$ (see text), for data in Figs.~\ref{subfig:IP_High_Int}--\ref{subfig:IP_Low_Int} represented by red (far-right), yellow and dark blue (far-left) curves respectively. The dashed lines indicate exponential fits used to determine $\theta_c$. The dot-dashed magenta (blue) curve indicates the single-electron PSL level for the BAS-SR (BAS-MS) plates used for Figs.~\ref{subfig:IP_High_Int} and \ref{subfig:IP_Med_Int} (Fig.~\ref{subfig:IP_Low_Int}) and the shaded regions are the estimated uncertainties in these levels based on measurements in \cite{bonnet_response_2013}. The solid data points represent our best estimate of $\theta_c$, and the associated uncertainty (see text).}
     \label{fig:IP_lineout_int}
\end{figure}

The primary goal of the analysis is to determine $\theta_c$. This can be achieved by obtaining an average line profile for varying $\theta$ that is representative of each image, to then find the smallest $\theta$ at which the signal falls to the PSL level that a single electron would deposit. First, to remove extraneous noise from the recorded image, a median filter was first applied in units of 3$\times$3 squares of 50 $\mu$m $\times$ 50 $\mu$m pixels. The data was then visually centered using the large-angle falloff in Figs.~\ref{subfig:IP_High_Int} and \ref{subfig:IP_Med_Int} with the crosshairs shown in white, which is accurate to $\pm$ 1 mm. As the BiAD in Fig.~\ref{subfig:IP_Low_Int} extends beyond the image plate, the center in this case was chosen in the vicinity of that in Figs.~\ref{subfig:IP_High_Int} and \ref{subfig:IP_Med_Int} relative to the hole cut in the plate, since the same alignment procedure was used for all three measurements. The uncertainty was extended to $\pm$ 5 mm for the measurement in Fig.~\ref{subfig:IP_Low_Int}. We define $g(\theta,\phi)$ as the PSL value stored in each pixel of the image, located at a given $\theta$ ($= \arctan{(\sqrt{x^2+y^2}/d)}$) and $\phi$ ($= \arctan{(y/x)}$) from the laser focus (as shown in Fig.~\ref{subfig:IP_schematic}). A $\phi$-averaged line profile, given by $h(\theta)= \int_{0}^{2\pi}g(\theta,\phi) d\phi\ /\int_{0}^{2\pi}d\phi$, for each image in Fig.~\ref{fig:IP_scans} is shown in Fig.~\ref{fig:IP_lineout_int}. We then obtain a $\theta_c$ that is representative of the recorded image by finding the smallest $\theta$ where $h(\theta) =$ the PSL value that a single electron in the MeV energy range would deposit on average in a single pixel on the image plate. This is estimated to be $\approx (9\pm 2) \times 10^{-3}$ PSL for BAS-SR and $\approx (3\pm 1) \times 10^{-2}$ PSL for BAS-MS from \cite{bonnet_response_2013} for normal incidence. To account for the incident angle dependence of the deposited PSL, this value was then multiplied by $1/\cos{(\theta})$ to obtain a signal floor as shown in Fig.~\ref{fig:IP_lineout_int}. The shaded region in magenta (pale blue) represents the signal floor for the BAS-SR (BAS-MS). The point of intersection of $h(\theta)$ and the corresponding floor was used as a first estimate to then apply an exponential fit to $h(\theta)$ from this point to an added $4^{\circ}$ (toward the left in Fig.~\ref{fig:IP_lineout_int}). We then found $\theta_c$ by extending the fit data to smaller $\theta$ to find where it crosses the shaded region, as shown in Fig.~\ref{fig:IP_lineout_int}. To account for the uncertainty in positioning the plate, the same procedure was repeated by randomly varying the position of the center and the distance of the plate from the focus within the limits of uncertainty detailed earlier. This dominates the contribution from the uncertainty in estimating the signal floor. The final $\theta_c$ value for each measurement is reported in Table \ref{tab:IP_params}. Additionally, to highlight the variation of the small-angle falloff position for different $\phi$, the same procedure for a fixed center was applied to individual radial line profiles, $g(\theta,\phi=\phi_0)$, with $\phi_0 \in [0^{\circ},359^{\circ}]$ in steps of $3^{\circ}$. These profiles were taken in strips (width 0.55 mm) to reduce fluctuations due to single-pixel noise. The position of the falloff point along each $\phi_0$ is shown by the white dot markers in Fig.~\ref{fig:IP_scans}.

\subsection{Inferring $a_0$} \label{ssec:a_0_estimation}

To compare the $\theta_c$ measured (Table \ref{tab:IP_params}) to that predicted by Eq.~(\ref{eqn:tan_theta_a_0}), we need an estimate of $a_0$ for each measurement. To that end, we infer $a_0$ from the pulse energy, $U$, delivered on the focal plane, the temporal pulse duration, $\tau$, (full width at half maximum) and images of the focal spot. Our procedure is described in detail in the appendix of \cite{he_towards_2019}. Briefly, after removing extraneous noise with a median filter in units of $3\times 3$ pixels of the recorded image (similar to that done in Sec.~\ref{ssec:IP_section}), the pulse energy ($U$) is distributed over the focal spot by scaling the peak signal in the image to $I = K U/\tau$. The proportionality constant, $K$, is determined by the focal-spot distribution. Specifically, $K = C_{pk}/(C_{sum} A_{pix})$ where $C_{pk}$ corresponds to the peak signal, $C_{sum}$ is the total signal and $A_{pix}$ is the area of the camera pixel in physical units ($\mathrm{cm^2}$, when $I$ is expressed in $\mathrm{W/cm^2}$). 

We measured the pulse energy in the full beam through a leaky mirror before the compressor on each shot. We then factored in losses suffered from the leaky-mirror transmittivity, the compressor, the reflectivity of the tuning and parabolic mirrors, and during beam transport. We report the average $U$ during each sequence of laser shots in Tables \ref{tab:IP_params} and \ref{tab:Sci_det_params} with an overall uncertainty estimated to be $\approx \pm 10\%$ after considering the systematic uncertainties in the calibration process. We measured $\tau$ for each laser shot by diverting a small portion of the beam from the target chamber into a second-harmonic autocorrelator placed outside a fused silica viewport (thickness 5 mm). The reported $\tau$ in Tables \ref{tab:IP_params} and \ref{tab:Sci_det_params} is the average value over each sequence of laser shots after correcting for the dispersion through the viewport and $\approx 1$ m of air in beam transport to the autocorrelator. We estimate that the overall uncertainty in the reported $\tau$ is $\pm 20\%$. The uncertainty in the precision of measuring the averages ($<2\%$) is dominated by the systematic uncertainties in measuring $U$ and $\tau$. A representative set of focal spot images taken over different days at low power, shown in Figs.~\ref{subfig:focal_spot}, \ref{subfig:focal_spot_2} and \ref{subfig:focal_spot_3}, indicate that there is some shot-to-shot fluctuation of the focal spot distribution as well as some consistent distortions. The fluctuation in the focal spot distribution affects the peak intensity for fixed $U$ and $\tau$ since it changes the area over which the pulse energy is distributed. Consequently, the peak intensity will decrease if there is an increase in the energy distributed away from the center of the focus. If $U = 23$ J, $\tau = 35$ fs, the peak intensity ($\propto U/\tau$) for Figs.~\ref{subfig:focal_spot}, \ref{subfig:focal_spot_2} and \ref{subfig:focal_spot_3} would be $\approx 2.0\times 10^{20}$, $1.9\times 10^{20}$ and $1.5\times 10^{20}$ W/cm$^2$ respectively, assuming that these images are representative of the focus at full power. Using this method, we observed that for fixed $U$ and $\tau$, the fluctuation in the focal spot distribution over different laser shots across different days contributes to an uncertainty $\approx 8\%$ in the intensity estimate. As a result, the overall uncertainty in the intensity derived from the focal spot method is estimated to be $\approx 24\%$. Therefore, we infer $a_0 \pm \delta a_0$ for the measurements in Figs.~\ref{subfig:IP_High_Int}, \ref{subfig:IP_Med_Int} and \ref{subfig:IP_Low_Int} to be $\approx 9.1\pm 1.1, 7.3\pm 0.9$ and $5.7\pm 0.7$ respectively.

\subsection{Scintillation electron detectors}\label{ssec:e_det}

\def\arraystretch{1.6}
\begin{table}
    \begin{center}
        \caption{Summary of experimental conditions used for measurements with the scintillation electron detectors. The $U$ and $\tau$ (as defined in Sec.~\ref{ssec:a_0_estimation}) reported here is the average value obtained over the corresponding sequence of shots, after considering the systematic uncertainties in the measurement process, which dominate the uncertainty in measuring the average. The techniques used to measure $U$ and $\tau$ are detailed in Sec.~\ref{ssec:a_0_estimation}.\\}
        \begin{tabular}{l l l l} 
            \hline\hline
            \multirow{2}{4em}{\textbf{Gas}} & \multirow{2}{5em}{\textbf{U (J) (\boldmath $\pm 10\%$)}} & \multirow{2}{5em}{\textbf{\boldmath $\tau$ (fs) (\boldmath $\pm 20\%$)}} & \multirow{2}{4.5em}{\textbf{Pressure (mbar)}}\\ \\
            \hline
            \multirow{2}{3em}{N$_2$} & 10.6 & 35.6 & $5\times10^{-5}$\\ 
            & 26.9 & 40.4 & $3\times10^{-5}$\\
            \multirow{2}{3em}{Ar} & 10.2 & 30.3 & $9\times10^{-5}$\\ 
            & 26.7 & 38.7 & $9\times10^{-5}$\\
            \multirow{2}{3em}{Xe} & 10.0 & 31.5 & $4\times10^{-5}$\\ 
            & 26.4 & 49.7 & $4\times10^{-5}$\\
            \hline\hline
        \end{tabular}
        \label{tab:Sci_det_params}
    \end{center}
\end{table}

\begin{figure}
     \centering
     \includegraphics[width=1\linewidth]{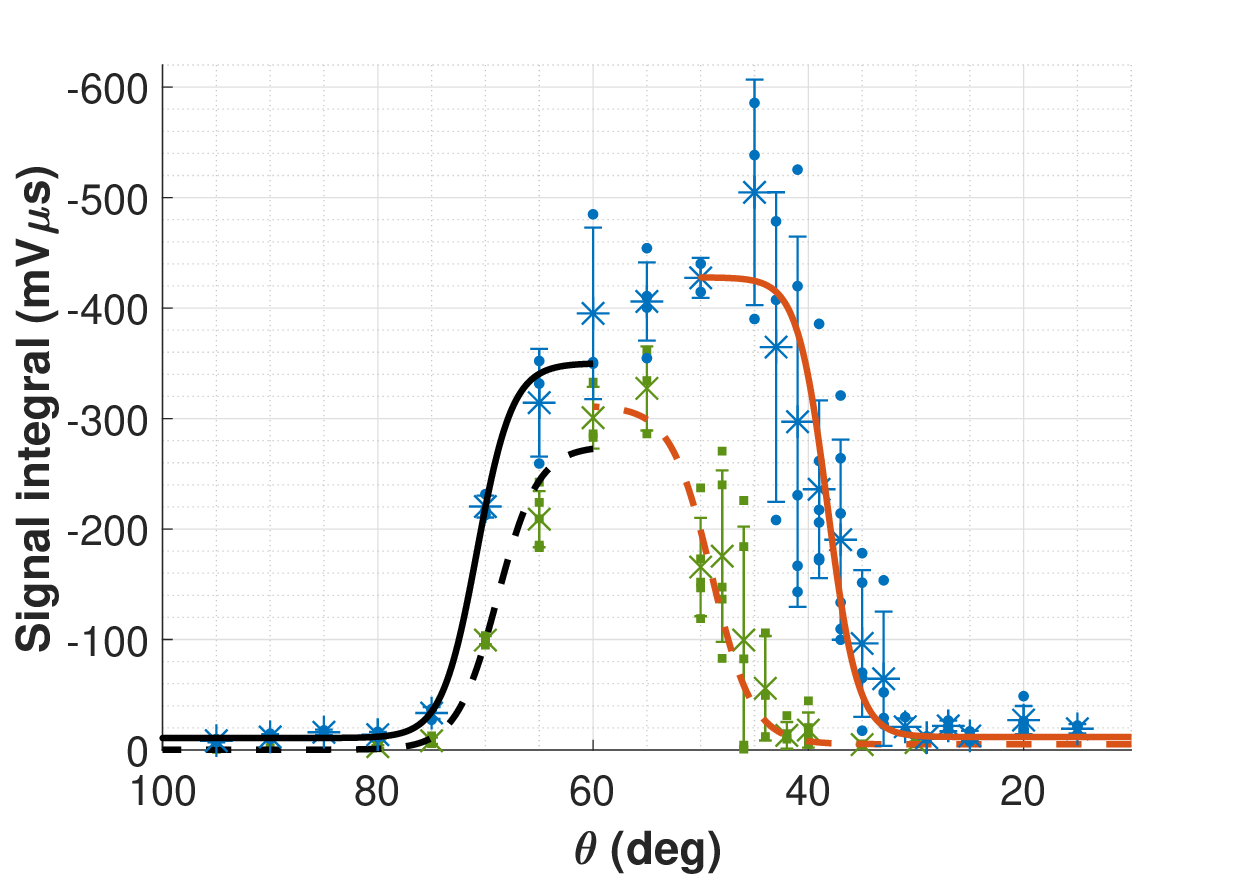}
     \caption{Single-shot scintillation data captured with Ar gas for $15^{\circ} \leq \theta \leq 95^{\circ}$ with 26.7 J (blue dots) and 10.2 J (green squares) of laser energy on target for the experimental conditions detailed in Table \ref{tab:Sci_det_params}. The average high (low) energy data are displayed as ``$*$" (``$\times$") along with their corresponding standard deviations. The solid (dashed) orange and black lines are sigmoid fits to the small and large angle falloff respectively for the high (low) energy measurement. The large angle threshold is set by the aluminum filters used to block the plethora of low energy electrons.}
     \label{fig:Scintillating_e_det_scans}
\end{figure}

While studying the average distribution over multiple shots reveals the average features of the laser focus, it is also important to characterize the shot-to-shot variation. We demonstrate proof of principle with the use of a scintillation electron detector that offers fast measurement capabilities, which may be extended to capture the BiAD of ejected electrons on each shot. Here, we observe shot-to-shot fluctuations for varying $\theta$ at a fixed distance from the focus in the $x$-$z$ plane. We give the experimental conditions for these measurements in Table \ref{tab:Sci_det_params}. The detector was comprised of a bismuth germanate (BGO) scintillator crystal  (thickness 3 mm), placed in front of a multi-pixel photon counter (MPPC) \cite{noauthor_mppcs_nodate} that was connected to an oscilloscope to record the electron signal. The front surface of the scintillator was located at 138 mm from the laser focus. A tungsten disk (thickness 3 mm) with a 3-mm hole was used as an aperture to limit the acceptance angle to $\approx 0.37$ msr. Two layers of Al foil (thickness 12 $\mu$m each), were used to block scattered laser light and the copious number of electrons with kinetic energy $\lessapprox 70$ keV. A layer of aluminized mylar (thickness 2 $\mu$m, with 100-nm Al on each side) was used to cover the front of the detector for added light tightness. The detector was placed on a rotatory stage as shown in Fig.~\ref{subfig:IP_schematic}. 

\begin{figure}
     \centering
     \includegraphics[width=1\linewidth]{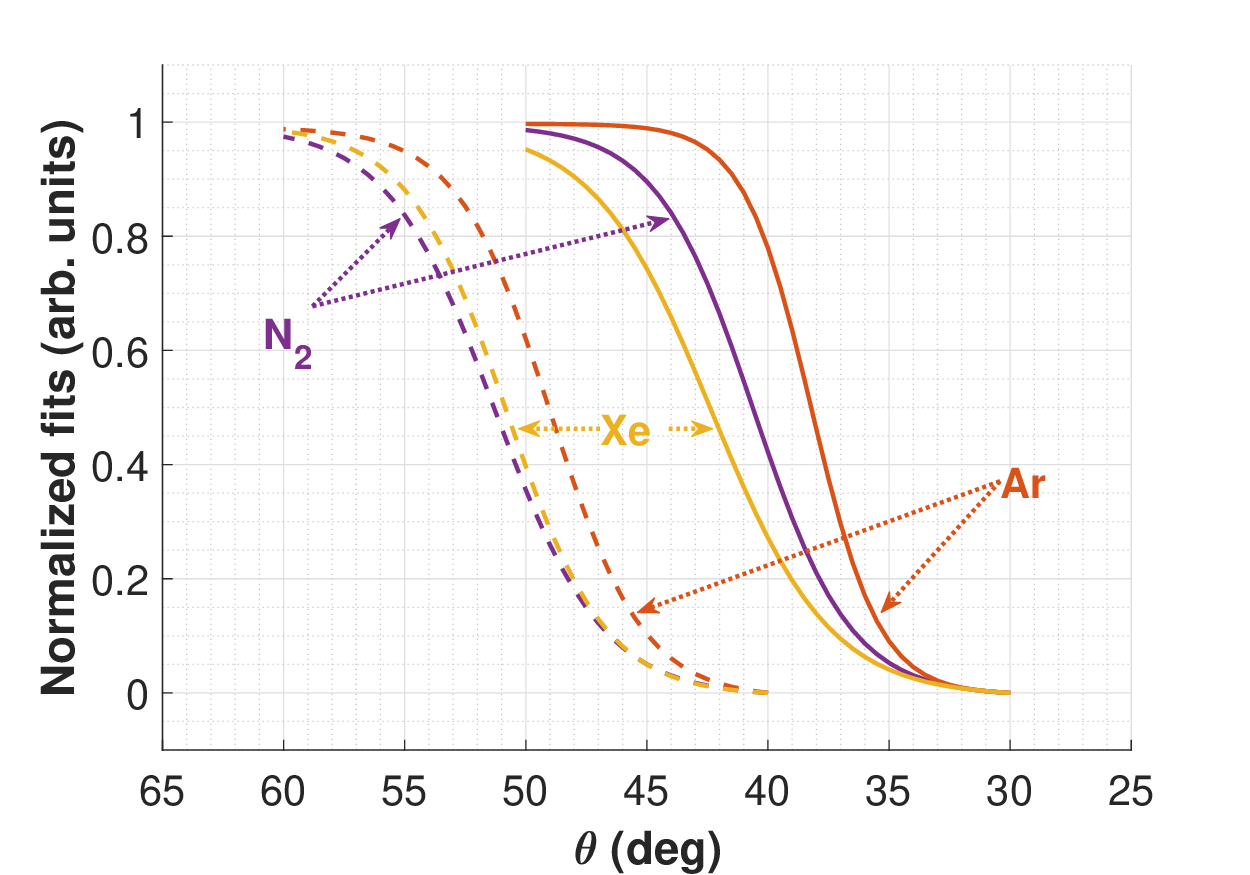}
     \caption{Normalized and background subtracted sigmoid fits for the small-angle falloff of measurements performed with N$_2$ (violet), Ar (orange) and Xe (yellow). The solid (dashed) curves correspond to high (low) energy measurements.}
     \label{fig:Overall_summary}
\end{figure}

The scintillation signal recorded by the MPPC, and then registered electronically on the oscilloscope, was integrated in time for each laser shot. This signal integral per shot was then plotted as a function of angular position of the detector as shown in Fig.~\ref{fig:Scintillating_e_det_scans}, which shows measurements made with Ar at low (10.2 J) and high (26.7 J) laser energy. By fitting a sigmoid $\theta > 60^{\circ}$ (shown in black), we show that the large-angle falloff is independent of the laser intensity and is caused by the Al foil that blocks electrons with energy $\lessapprox$ 70 keV \cite{noauthor_stopping_nodate}. This is evident from Fig.~\ref{fig:Scintillating_e_det_scans} where the black solid and dashed lines fall to $5\%$ at almost the same $\theta$ ($\approx 75^{\circ}$). This large-angle falloff at $\theta \approx$ 75$^{\circ}$ is also consistent with the predicted energy from Eq.~(\ref{eqn:tan_theta}), as similarly discussed for the large-angle falloff of the image plate measurements in Sec.~\ref{ssec:IP_section}. To highlight the shift in the small-angle falloff, a sigmoid curve was fit in the range $\theta < 60^{\circ}$ and $\theta < 50^{\circ}$ for the low and high energy cases respectively. A summary of all the sigmoid fits for the small-angle falloff can be seen in Fig.~\ref{fig:Overall_summary}, which is consistent with the results from Sec.~\ref{ssec:IP_section} in that the electrons are ejected at smaller angles for higher peak intensities.

\section{Discussion} \label{sec:discussion_section}

\begin{figure}
     \centering
     \includegraphics[width=1\linewidth]{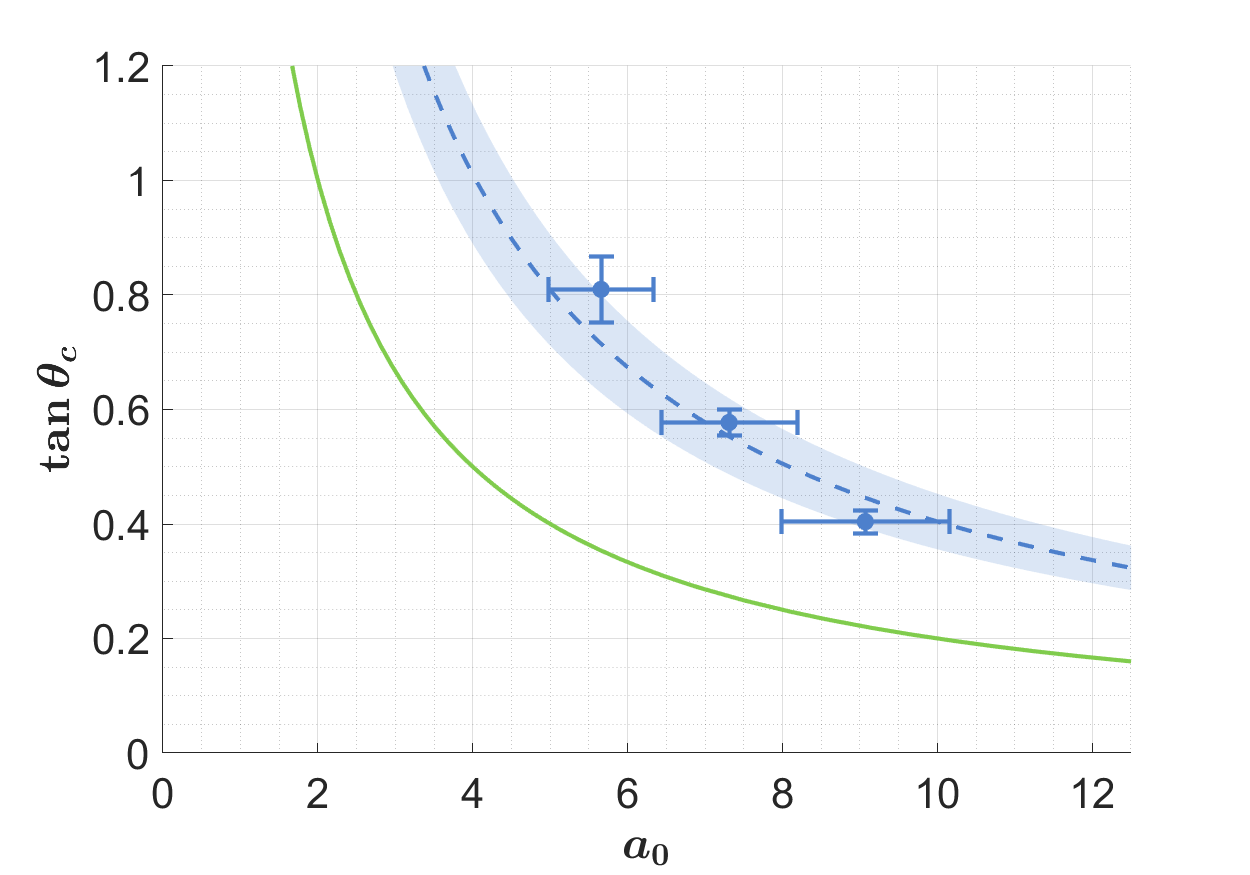}
     \caption{Energy scaling, $\tan{\theta_c}$ vs inferred $a_0$ -- $9.1\pm 1.1$, $7.3\pm 0.9$ and $5.7\pm 0.7$ for the three measurements in Fig.~\ref{fig:IP_scans} shown by blue filled circles, with Eq.~(\ref{eqn:tan_theta_a_0}) shown by the green solid curve. The inferred $a_0$ is based on the conditions given in Table \ref{tab:IP_params} (see Sec.~\ref{ssec:a_0_estimation}). The horizontal error bars, $\pm \delta a_0$, represent the uncertainty in calculating $a_0$ using the low-power focal spot image (Figs.~\ref{subfig:focal_spot}-\ref{subfig:focal_spot_3}). The vertical error bars represent the uncertainty in the measurement of $\tan{\theta_c}$ (see Sec.~\ref{ssec:IP_section}). The blue dashed line represents a numerical fit of the data to $\tan{\theta_c} = 2\eta/a_0$, where $\eta = 2.02^{+0.26}_{-0.22}$; the uncertainty in $\eta$ is represented by the shaded region around the dashed curve.}
     \label{fig:Scaling_plot}
\end{figure}

From the results presented in Fig.~\ref{fig:IP_lineout_int}, it is clear that $\theta_c$ scales with the laser energy and pulse duration, and presumably intensity. We now compare this scaling with that predicted by Eq.~(\ref{eqn:tan_theta_a_0}) using the $a_0$ estimates from Sec.~\ref{ssec:a_0_estimation}. Figure \ref{fig:Scaling_plot} compares the measured and theoretical scaling of $\tan{\theta_c}$ with $a_0$ for the three image plate measurements in Figs.~\ref{subfig:IP_High_Int}, \ref{subfig:IP_Med_Int} and \ref{subfig:IP_Low_Int} with $a_0 \pm \delta a_0 \approx 9.1 \pm 1.1$, $7.3 \pm 0.9$ and $5.7 \pm 0.7$ respectively. We first point out that the data is statistically consistent with $\tan{\theta_c} \propto 1/a_0$. It is also evident that the data is not consistent with the coefficient of 2. Fitting the experimental data to the function $\tan{\theta_c} = 2\eta/a_x$, to the measured data at $a_x=a_0$ (shown by the blue dashed line), $a_x = a_0 + \delta a_0$ and $a_x = a_0 - \delta a_0$ (represented by the boundaries of the shaded region), $\eta$ was found to be $2.02^{+0.26}_{-0.22}$. 

There could be several reasons why $\eta$ differs from 1: (i) the inferred $a_0$ might not correspond to the peak intensity present in the focus, (ii) the measured $\theta_c$ may not reflect the peak intensity in the focus (iii) the need for a more robust theoretical model of the dynamics in a focused laser pulse. 

Here, we discuss some factors that could likely contribute to (i). First, the focal spot may not behave the same way at high power as it did at low power, and might possibly expand in size or have a larger proportion of energy distributed farther away from the central spot. For example, it is straightforward to show that for a Gaussian beam, since $a_0 \propto 1/w_0$, the beam waist at full power would only have to expand to twice of that at low power to reduce the full power estimate of $a_0$ by half, which would then be a closer fit to the plane-wave model. Second, if there were a pulse front tilt, the effective pulse duration would be longer. For example, Fig.~1 of \cite{ouatu_ionization_2022} shows that a tilt of only 0.2 fs/mm for a beam of 25-cm diameter (as in the VEGA-3 laser) results in a reduction of the intensity by a factor of $\approx 3.2$. Additionally, other aberrations such as astigmatism or coma that might cause the focal spot to be oval-like or teardrop-like (such as in the central region of Figs.~\ref{subfig:focal_spot}--\ref{subfig:focal_spot_3}), could also contribute to lowering the intensity delivered in the focus.  

To address the possibility of (ii), we note that we measured $\theta_c$ by finding the smallest $\theta$ at which the $\phi$-averaged pixel response over $\approx$ 100 laser shots meets the PSL level that a single electron would deposit on average in a pixel from \cite{bonnet_response_2013}. This intersection was found to be well above the noise (see Fig.~\ref{fig:IP_lineout_int}), implying that the image plate was irradiated by more than one electron on average (for all $\phi$) for $\theta > \theta_c$. However, if one were to perform the measurement with more laser shots, it is possible that the $\phi$-averaged pixel response could reach the single-electron PSL level at a $\theta < \theta_c$. Therefore, the measured $\theta_c$ here may be used as a reasonable upper bound for the smallest ejection angle of electrons that presumably reflects the peak intensity.

A detailed simulation and additional experiments are required to address (iii), which is an ongoing project.  

Another important aspect of the image plate measurements in Fig.~\ref{fig:IP_scans} is the asymmetry in the ring-like distribution of ejected electrons around $\vec{k}$, in that there is a higher concentration of electrons on one side of each image. This asymmetry could be explained by the existence of a pulse front tilt causing a preferential sweeping of the electrons in one direction by the part that arrives earlier, creating a relative paucity of electrons on the other side. Ponderomotive ejection of electrons to one side has been reported for beams with larger phase-front tilts \cite{wilhelm_tilted_2019}. Further, the noticeable similarities in the distortions among the images in Fig.~\ref{fig:IP_scans}, coupled with the fact that each image is a result of $\approx$ 100 shots on different days, suggest that these distortions may be related to some consistent distortions in the beam focus. Consequently, it is plausible that detailed information about the distortions in the laser focus at full power is now accessible by measuring the BiAD of ejected electrons that, to the best of our knowledge, has been observed for the first time for all $\phi$. We expect that studying the factors that influence the distortions in the BiAD of ejected electrons may enable a way to identify and correct for the distortions in the focus at full power. This is being investigated further with simulations modeling the interaction of the electrons with laser beams having different aberrations, as well as planned experiments.  

While this manuscript explores the scaling of $\tan{\theta_c}$ with $a_0$ inferred from measurements made at low power, it may be possible to enable the direct measurement of the peak intensity by measuring $\theta_c$. This requires gaining a better understanding of the relationship between $\gamma_p$ and $a_0$, possibly by studying the change in $\gamma_p$ for different conditions at these intensities through future experiments to test the applicability of Eq.~(\ref{eqn:tan_theta_a_0}). 

Using $\theta_c$ alone to measure $a_0$ does have the limitation posed by the asymptotic behavior of $\tan{\theta_c} \propto 1/a_0$ that causes the relative uncertainty in $a_0$, $ \Delta a_0/a_0 \approx 2\delta \theta_c/\sin{2\theta_c}$, to increase rapidly as $\theta_c \to 0^{\circ}$. For example, if one wishes to measure $a_0$ ($I$) to within $\pm 12.5\%$ ($\pm 25\%$) with a fixed precision in measuring $\theta_c$, $\delta \theta_c = \pm 1^{\circ}$, we can show that this method may only be usable for $\theta_c \gtrapprox 8^{\circ}$. In other words, with the assumption of a parameterized plane-wave model, $\tan{\theta} = 2\eta/a_0$, this method starts to be less effective beyond  $I \approx 4\eta^2\times 10^{20}\ \mathrm{W/cm^2}$ as the relative uncertainty, $\Delta I/I$, increases rapidly. If $\eta = 2$, as measured in this experiment, this restricts the applicability of this technique to $I \lessapprox 10^{21} \ \mathrm{W/cm^2}$. 

Therefore, it will be important to measure $\gamma_p$ over all $\phi$ outside this intensity range. It is critical to test Eq.~(\ref{eqn:tan_theta}) for smaller f/\# by comparing $\theta_c$ with the ejection angle of electrons with $\gamma = \gamma_p$, to explore the limitations of applying the paraxial model to study the interaction. For these tighter focusing geometries, the substantial spread in the direction of $\vec{k}$ before and after focus may play a significant role in the scattering angle of the most energetic electrons. One may also study the onset of collective effects at higher gas pressures to estimate the range in which the electrons may be treated as single particles. Further investigations are underway to understand these effects in greater detail and to test the applicability of this approach for such cases.

\section{Conclusion}

Through this study, we have provided a technique to assess the peak intensity of petawatt-class lasers at full power by measuring $\theta_c$ and highlighted the need to study Eqs.~(\ref{eqn:tan_theta}) and (\ref{eqn:gamma_hartemann}) in greater detail. We have shown that the use of image plates can be an exceedingly straightforward, yet powerful, method to measure $\theta_{c}$ under different experimental conditions and to capture distortions in the BiAD of ejected electrons. Using the image plate measurements, we compared our experimental data to the plane-wave parameterization in Eq.~(\ref{eqn:tan_theta_a_0}) to find that our data, although in agreement with $\tan{\theta_c} \propto 1/a_0$, closely fits $\tan{\theta} = 2\eta/a_0$, differing from Eq.~(\ref{eqn:tan_theta_a_0}) by $\eta = 2.02^{+0.26}_{-0.22}$. We discussed possible factors that could contribute to a lower peak intensity in the laser focus compared to the estimate from low-power measurements, which would then be in better agreement with the theory. We also discussed the sensitivity of the BiAD of ejected electrons to the realistic non-idealities of a high intensity laser focus that can cause significant deviation from ideal Gaussian behavior. We have demonstrated the use of a scintillation electron detector to observe shot-to-shot fluctuations, as seen in Fig.~\ref{fig:Scintillating_e_det_scans}, and note that it may also be extended to measure $\theta_c$ with higher sampling. A modification of this principle may also enable the measurement of the BiAD of ejected electrons over all $\phi$ on each shot, allowing real-time monitoring of the peak intensity, and possibly beam distortions, in the focal spot at full power. While we demonstrated the use of our instrumentation for intensities in the range of $10^{19}$ to $10^{20}$ W/cm$^2$, we believe that these techniques could be applied to the broader range of intensities from $\approx 10^{18}$ to $10^{21}$ W/cm$^2$. This method may also hold promise at higher intensities when the acceleration of protons is expected to become relativistic in a single cycle. The techniques presented herein should aid users of high powered laser facilities across the world and across disciplines in assessing the focal volume of petawatt-class lasers while studying numerous phenomena that occur in this intensity regime.

\begin{acknowledgments}
The authors would like to thank Scott Wilks for helpful discussions regarding the electron energy dependence on the laser intensity, Lester Putnam for technical help in preparation for the experiment and everyone at CLPU for their assistance toward the smooth execution of the experiment. This work was supported by the National Science Foundation (PHY2010392), Natural Sciences and Engineering Research Council of Canada (RGPIN-2019-05013), Junta de Castilla y Le\'on (UIC-167) CLP087U16, Ministerio de Ciencia e Innovacion (PALMA Grant FIS2016-81056-R), LaserLab Europe V (GA871124). SR acknowledges support from the Kulkarni Graduate Student Summer Research Fellowship.   
\end{acknowledgments}


\bibliography{IP_ref_paper_bib_no_url,IP_new_physics_bib_no_url,IP_books_bib,IP_ref_articles_bib}

\begin{thebibliography}{52}
\providecommand{\natexlab}[1]{#1}
\providecommand{\url}[1]{\texttt{#1}}
\expandafter\ifx\csname urlstyle\endcsname\relax
  \providecommand{\doi}[1]{doi: #1}\else
  \providecommand{\doi}{doi: \begingroup \urlstyle{rm}\Url}\fi

\bibitem[Danson et~al.(2015)Danson, Hillier, Hopps, and
  Neely]{danson_petawatt_2015}
Colin Danson, David Hillier, Nicholas Hopps, and David Neely.
\newblock Petawatt class lasers worldwide.
\newblock \emph{High Power Laser Science and Engineering}, 3:\penalty0 e3,
  2015.
\newblock ISSN 2095-4719, 2052-3289.
\newblock \doi{10.1017/hpl.2014.52}.
\newblock Publisher: Cambridge University Press.

\bibitem[Yoon et~al.(2021)Yoon, Kim, Choi, Sung, Lee, Lee, and
  Nam]{yoon_realization_2021}
Jin~Woo Yoon, Yeong~Gyu Kim, Il~Woo Choi, Jae~Hee Sung, Hwang~Woon Lee,
  Seong~Ku Lee, and Chang~Hee Nam.
\newblock Realization of laser intensity over 10 $^{\textrm{23}}$ {W}/cm
  $^{\textrm{2}}$.
\newblock \emph{Optica}, 8\penalty0 (5):\penalty0 630, May 2021.
\newblock ISSN 2334-2536.
\newblock \doi{10.1364/OPTICA.420520}.

\bibitem[Yoon et~al.(2022)Yoon, Sung, Lee, Lee, and Nam]{yoon_ultra-high_2022}
Jin~Woo Yoon, Jae~Hee Sung, Seong~Ku Lee, Hwang~Woon Lee, and Chang~Hee Nam.
\newblock Ultra-high intensity lasers as tools for novel physics.
\newblock \emph{Journal of the Korean Physical Society}, 81\penalty0
  (6):\penalty0 562--569, September 2022.
\newblock ISSN 1976-8524.
\newblock \doi{10.1007/s40042-022-00411-3}.

\bibitem[Piazza et~al.()Piazza, Willingale, Zuegel, Areﬁev, Banerjee,
  Blackburn, Bulanov, Gregori, Howell, Park, Roth, Darmstadt, Shvets, Spohr,
  Tanaka, Uzdensky, Wilks, and Zurek]{piazza_multi-petawatt_nodate}
Antonino~Di Piazza, Louise Willingale, Jonathan Zuegel, Alexey Areﬁev, Sudeep
  Banerjee, Tom Blackburn, Stepan Bulanov, Gianluca Gregori, Calvin Howell,
  Hye-Sook Park, Markus Roth, TU~Darmstadt, Gennady Shvets, Klaus Spohr, Kazuo
  Tanaka, Dmitri Uzdensky, Scott Wilks, and Eva Zurek.
\newblock Multi-{Petawatt} {Physics} {Prioritization} {Workshop} {Report}.

\bibitem[Bell and Kirk(2008)]{bell_possibility_2008}
A.~R. Bell and John~G. Kirk.
\newblock Possibility of {Prolific} {Pair} {Production} with {High}-{Power}
  {Lasers}.
\newblock \emph{Physical Review Letters}, 101\penalty0 (20):\penalty0 200403,
  November 2008.
\newblock \doi{10.1103/PhysRevLett.101.200403}.
\newblock Publisher: American Physical Society.

\bibitem[Krajewska et~al.(2013)Krajewska, Müller, and
  Kamiński]{krajewska_bethe-heitler_2013}
K.~Krajewska, C.~Müller, and J.~Z. Kamiński.
\newblock Bethe-{Heitler} pair production in ultrastrong short laser pulses.
\newblock \emph{Physical Review A}, 87\penalty0 (6):\penalty0 062107, June
  2013.
\newblock \doi{10.1103/PhysRevA.87.062107}.
\newblock Publisher: American Physical Society.

\bibitem[Zhu et~al.(2016)Zhu, Yu, Sheng, Yin, Turcu, and
  Pukhov]{zhu_dense_2016}
Xing-Long Zhu, Tong-Pu Yu, Zheng-Ming Sheng, Yan Yin, Ion Cristian~Edmond
  Turcu, and Alexander Pukhov.
\newblock Dense {GeV} electron–positron pairs generated by lasers in
  near-critical-density plasmas.
\newblock \emph{Nature Communications}, 7\penalty0 (1):\penalty0 13686,
  December 2016.
\newblock ISSN 2041-1723.
\newblock \doi{10.1038/ncomms13686}.
\newblock Number: 1 Publisher: Nature Publishing Group.

\bibitem[Salgado et~al.(2021)Salgado, Grafenstein, Golub, Döpp, Eckey,
  Hollatz, Müller, Seidel, Seipt, Karsch, and Zepf]{salgado_towards_2021}
F.~C. Salgado, K.~Grafenstein, A.~Golub, A.~Döpp, A.~Eckey, D.~Hollatz,
  C.~Müller, A.~Seidel, D.~Seipt, S.~Karsch, and M.~Zepf.
\newblock Towards pair production in the non-perturbative regime.
\newblock \emph{New Journal of Physics}, 23\penalty0 (10):\penalty0 105002,
  October 2021.
\newblock ISSN 1367-2630.
\newblock \doi{10.1088/1367-2630/ac2921}.
\newblock Publisher: IOP Publishing.

\bibitem[Karplus and Neuman(1950)]{karplus_non-linear_1950}
Robert Karplus and Maurice Neuman.
\newblock Non-{Linear} {Interactions} between {Electromagnetic} {Fields}.
\newblock \emph{Physical Review}, 80\penalty0 (3):\penalty0 380--385, November
  1950.
\newblock ISSN 0031-899X.
\newblock \doi{10.1103/PhysRev.80.380}.

\bibitem[Costantini et~al.(1971)Costantini, De~Tollis, and
  Pistoni]{costantini_nonlinear_1971}
V.~Costantini, B.~De~Tollis, and G.~Pistoni.
\newblock Nonlinear effects in quantum electrodynamics.
\newblock \emph{Il Nuovo Cimento A (1965-1970)}, 2\penalty0 (3):\penalty0
  733--787, April 1971.
\newblock ISSN 1826-9869.
\newblock \doi{10.1007/BF02736745}.

\bibitem[Gies et~al.(2018)Gies, Karbstein, Kohlfürst, and
  Seegert]{gies_photon-photon_2018}
Holger Gies, Felix Karbstein, Christian Kohlfürst, and Nico Seegert.
\newblock Photon-photon scattering at the high-intensity frontier.
\newblock \emph{Physical Review D}, 97\penalty0 (7):\penalty0 076002, April
  2018.
\newblock \doi{10.1103/PhysRevD.97.076002}.
\newblock Publisher: American Physical Society.

\bibitem[Roso et~al.(2022)Roso, Lera, Ravichandran, Longman, He,
  Pérez-Hernández, Apiñaniz, Smith, Fedosejevs, and Hill]{roso_towards_2022}
Luis Roso, Roberto Lera, Smrithan Ravichandran, Andrew Longman, Calvin~Z. He,
  José~Antonio Pérez-Hernández, Jon~I. Apiñaniz, Lucas~D. Smith, Robert
  Fedosejevs, and Wendell~T. Hill.
\newblock Towards a direct measurement of the quantum-vacuum {Lagrangian}
  coupling coefficients using two counterpropagating super-intense laser
  pulses.
\newblock \emph{New Journal of Physics}, 24\penalty0 (2):\penalty0 025010,
  March 2022.
\newblock ISSN 1367-2630.
\newblock \doi{10.1088/1367-2630/ac51a7}.
\newblock Publisher: IOP Publishing.

\bibitem[noa({\natexlab{a}})]{noauthor_zeus_nodate}
{ZEUS}, {\natexlab{a}}.
\newblock URL \url{https://zeus.engin.umich.edu//}.

\bibitem[noa({\natexlab{b}})]{noauthor_clf_nodate}
{CLF} {Vulcan} 2020 {Upgrade}, {\natexlab{b}}.
\newblock URL \url{https://www.clf.stfc.ac.uk/Pages/Vulcan-2020.aspx}.

\bibitem[Bromage et~al.(2019)Bromage, Bahk, Begishev, Dorrer, Guardalben,
  Hoffman, Oliver, Roides, Schiesser, Iii, Spilatro, Webb, Weiner, and
  Zuegel]{bromage_technology_2019}
J.~Bromage, S.-W. Bahk, I.~A. Begishev, C.~Dorrer, M.~J. Guardalben, B.~N.
  Hoffman, J.~B. Oliver, R.~G. Roides, E.~M. Schiesser, M.~J.~Shoup Iii,
  M.~Spilatro, B.~Webb, D.~Weiner, and J.~D. Zuegel.
\newblock Technology development for ultraintense all-{OPCPA} systems.
\newblock \emph{High Power Laser Science and Engineering}, 7:\penalty0 e4,
  2019.
\newblock ISSN 2095-4719, 2052-3289.
\newblock \doi{10.1017/hpl.2018.64}.
\newblock Publisher: Cambridge University Press.

\bibitem[noa({\natexlab{c}})]{noauthor_apollon_nodate}
Apollon {Facility}, {\natexlab{c}}.
\newblock URL \url{https://apollonlaserfacility.cnrs.fr/en/facility/}.

\bibitem[Radier et~al.(2022)Radier, Chalus, Charbonneau, Thambirajah,
  Deschamps, David, Barbe, Etter, Matras, Ricaud, Leroux, Richard, Lureau,
  Baleanu, Banici, Gradinariu, Caldararu, Capiteanu, Naziru, Diaconescu, Iancu,
  Dabu, Ursescu, Dancus, Ur, Tanaka, and Zamfir]{radier_10_2022}
Christophe Radier, Olivier Chalus, Mathilde Charbonneau, Shanjuhan Thambirajah,
  Guillaume Deschamps, Stephane David, Julien Barbe, Eric Etter, Guillaume
  Matras, Sandrine Ricaud, Vincent Leroux, Caroline Richard, François Lureau,
  Andrei Baleanu, Romeo Banici, Andrei Gradinariu, Constantin Caldararu,
  Cristian Capiteanu, Andrei Naziru, Bogdan Diaconescu, Vicentiu Iancu, Razvan
  Dabu, Daniel Ursescu, Ioan Dancus, Calin~Alexandru Ur, Kazuo~A. Tanaka, and
  Nicolae~Victor Zamfir.
\newblock 10 {PW} peak power femtosecond laser pulses at {ELI}-{NP}.
\newblock \emph{High Power Laser Science and Engineering}, 10:\penalty0 e21,
  2022.
\newblock ISSN 2095-4719, 2052-3289.
\newblock \doi{10.1017/hpl.2022.11}.
\newblock Publisher: Cambridge University Press.

\bibitem[{Peng Yujie} et~al.(2021){Peng Yujie}, {Xu Yi}, and {Yu
  Lianghong}]{peng_yujie_overview_2021}
{Peng Yujie}, {Xu Yi}, and {Yu Lianghong}.
\newblock Overview and status of {Station} of {Extreme} {Light} toward 100
  {PW}.
\newblock \emph{Reza Kenkyu}, 49\penalty0 (2):\penalty0 93--96, 2021.
\newblock ISSN 0387-0200.
\newblock Place: Japan INIS Reference Number: 52108302.

\bibitem[Akturk et~al.(2010)Akturk, Gu, Bowlan, and
  Trebino]{akturk_spatio-temporal_2010}
Selcuk Akturk, Xun Gu, Pamela Bowlan, and Rick Trebino.
\newblock Spatio-temporal couplings in ultrashort laser pulses.
\newblock \emph{Journal of Optics}, 12\penalty0 (9):\penalty0 093001, August
  2010.
\newblock ISSN 2040-8986.
\newblock \doi{10.1088/2040-8978/12/9/093001}.

\bibitem[Ouatu et~al.(2022)Ouatu, Spiers, Aboushelbaya, Feng, von~der Leyen,
  Paddock, Timmis, Ticos, Krushelnick, and Norreys]{ouatu_ionization_2022}
I.~Ouatu, B.~T. Spiers, R.~Aboushelbaya, Q.~Feng, M.~W. von~der Leyen, R.~W.
  Paddock, R.~Timmis, C.~Ticos, K.~M. Krushelnick, and P.~A. Norreys.
\newblock Ionization states for the multipetawatt laser-{QED} regime.
\newblock \emph{Physical Review E}, 106\penalty0 (1):\penalty0 015205, July
  2022.
\newblock \doi{10.1103/PhysRevE.106.015205}.
\newblock Publisher: American Physical Society.

\bibitem[Harvey(2018)]{harvey_situ_2018}
C.~N. Harvey.
\newblock \textit{In situ} characterization of ultraintense laser pulses.
\newblock \emph{Physical Review Accelerators and Beams}, 21\penalty0
  (11):\penalty0 114001, November 2018.
\newblock ISSN 2469-9888.
\newblock \doi{10.1103/PhysRevAccelBeams.21.114001}.

\bibitem[Gao(2006)]{gao_laser_2006}
Ju~Gao.
\newblock Laser intensity measurement by {Thomson} scattering.
\newblock \emph{Applied Physics Letters}, 88\penalty0 (9):\penalty0 091105,
  February 2006.
\newblock ISSN 0003-6951.
\newblock \doi{10.1063/1.2180869}.
\newblock Publisher: American Institute of Physics.

\bibitem[Har-Shemesh and Di~Piazza(2012)]{har-shemesh_peak_2012}
Omri Har-Shemesh and Antonino Di~Piazza.
\newblock Peak intensity measurement of relativistic lasers via nonlinear
  {Thomson} scattering.
\newblock \emph{Optics Letters}, 37\penalty0 (8):\penalty0 1352, April 2012.
\newblock ISSN 0146-9592, 1539-4794.
\newblock \doi{10.1364/OL.37.001352}.

\bibitem[He et~al.(2019)He, Longman, Pérez-Hernández, de~Marco, Salgado,
  Zeraouli, Gatti, Roso, Fedosejevs, and Hill]{he_towards_2019}
C.~Z. He, A.~Longman, J.~A. Pérez-Hernández, M.~de~Marco, C.~Salgado,
  G.~Zeraouli, G.~Gatti, L.~Roso, R.~Fedosejevs, and W.~T. Hill.
\newblock Towards an in situ, full-power gauge of the focal-volume intensity of
  petawatt-class lasers.
\newblock \emph{Optics Express}, 27\penalty0 (21):\penalty0 30020, October
  2019.
\newblock ISSN 1094-4087.
\newblock \doi{10.1364/OE.27.030020}.

\bibitem[Link et~al.(2006)Link, Chowdhury, Morrison, Ovchinnikov, Offermann,
  Van~Woerkom, Freeman, Pasley, Shipton, Beg, Rambo, Schwarz, Geissel, Edens,
  and Porter]{link_development_2006}
Anthony Link, Enam~A. Chowdhury, John~T. Morrison, Vladimir~M. Ovchinnikov,
  Dustin Offermann, Linn Van~Woerkom, Richard~R. Freeman, John Pasley, Erik
  Shipton, Farhat Beg, Patrick Rambo, Jens Schwarz, Matthias Geissel, Aaron
  Edens, and John~L. Porter.
\newblock Development of an in situ peak intensity measurement method for
  ultraintense single shot laser-plasma experiments at the {Sandia} {Z}
  petawatt facility.
\newblock \emph{Review of Scientific Instruments}, 77\penalty0 (10):\penalty0
  10E723, October 2006.
\newblock ISSN 0034-6748.
\newblock \doi{10.1063/1.2336469}.
\newblock Publisher: American Institute of Physics.

\bibitem[Ciappina et~al.(2019)Ciappina, Popruzhenko, Bulanov, Ditmire, Korn,
  and Weber]{ciappina_progress_2019}
M.~F. Ciappina, S.~V. Popruzhenko, S.~V. Bulanov, T.~Ditmire, G.~Korn, and
  S.~Weber.
\newblock Progress toward atomic diagnostics of ultrahigh laser intensities.
\newblock \emph{Physical Review A}, 99\penalty0 (4):\penalty0 043405, April
  2019.
\newblock ISSN 2469-9926, 2469-9934.
\newblock \doi{10.1103/PhysRevA.99.043405}.

\bibitem[Galkin et~al.(2010)Galkin, Kalashnikov, Klinkov, Korobkin, Romanovsky,
  and Shiryaev]{galkin_electrodynamics_2010}
A.~L. Galkin, M.~P. Kalashnikov, V.~K. Klinkov, V.~V. Korobkin, M.~Yu.
  Romanovsky, and O.~B. Shiryaev.
\newblock Electrodynamics of electron in a superintense laser field: {New}
  principles of diagnostics of relativistic laser intensity.
\newblock \emph{Physics of Plasmas}, 17\penalty0 (5):\penalty0 053105, May
  2010.
\newblock ISSN 1070-664X, 1089-7674.
\newblock \doi{10.1063/1.3425864}.

\bibitem[Kalashnikov et~al.(2015)Kalashnikov, Andreev, Ivanov, Galkin,
  Korobkin, Romanovsky, Shiryaev, Schnuerer, Braenzel, and
  Trofimov]{kalashnikov_diagnostics_2015}
M.~Kalashnikov, A.~Andreev, K.~Ivanov, A.~Galkin, V.~Korobkin, M.~Romanovsky,
  O.~Shiryaev, M.~Schnuerer, J.~Braenzel, and V.~Trofimov.
\newblock Diagnostics of peak laser intensity based on the measurement of
  energy of electrons emitted from laser focal region.
\newblock \emph{Laser and Particle Beams}, 33\penalty0 (3):\penalty0 361--366,
  September 2015.
\newblock ISSN 0263-0346, 1469-803X.
\newblock \doi{10.1017/S0263034615000403}.

\bibitem[Ivanov et~al.(2018)Ivanov, Tsymbalov, Vais, Bochkarev, Volkov,
  Bychenkov, and Savel'ev]{ivanov_accelerated_2018}
K.~A. Ivanov, I.~N. Tsymbalov, O.~E. Vais, S.~G. Bochkarev, R.~V. Volkov, V.~Yu
  Bychenkov, and A.~B. Savel'ev.
\newblock Accelerated electrons for in situ peak intensity monitoring of
  tightly focused femtosecond laser radiation at high intensities.
\newblock \emph{Plasma Physics and Controlled Fusion}, 60\penalty0
  (10):\penalty0 105011, September 2018.
\newblock ISSN 0741-3335.
\newblock \doi{10.1088/1361-6587/aada60}.
\newblock Publisher: IOP Publishing.

\bibitem[Mackenroth et~al.(2019)Mackenroth, Holkundkar, and
  Schlenvoigt]{mackenroth_ultra-intense_2019}
Felix Mackenroth, Amol~R Holkundkar, and Hans-Peter Schlenvoigt.
\newblock Ultra-intense laser pulse characterization using ponderomotive
  electron scattering.
\newblock \emph{New Journal of Physics}, 21\penalty0 (12):\penalty0 123028,
  December 2019.
\newblock ISSN 1367-2630.
\newblock \doi{10.1088/1367-2630/ab5c4d}.

\bibitem[Hartemann et~al.(1995)Hartemann, Fochs, Le~Sage, Luhmann, Woodworth,
  Perry, Chen, and Kerman]{hartemann_nonlinear_1995}
F.~V. Hartemann, S.~N. Fochs, G.~P. Le~Sage, N.~C. Luhmann, J.~G. Woodworth,
  M.~D. Perry, Y.~J. Chen, and A.~K. Kerman.
\newblock Nonlinear ponderomotive scattering of relativistic electrons by an
  intense laser field at focus.
\newblock \emph{Physical Review E}, 51\penalty0 (5):\penalty0 4833--4843, May
  1995.
\newblock ISSN 1063-651X, 1095-3787.
\newblock \doi{10.1103/PhysRevE.51.4833}.

\bibitem[Quesnel and Mora(1998)]{quesnel_theory_1998}
Brice Quesnel and Patrick Mora.
\newblock Theory and simulation of the interaction of ultraintense laser pulses
  with electrons in vacuum.
\newblock \emph{Physical Review E}, 58\penalty0 (3):\penalty0 3719--3732,
  September 1998.
\newblock ISSN 1063-651X, 1095-3787.
\newblock \doi{10.1103/PhysRevE.58.3719}.

\bibitem[Moore et~al.(1995)Moore, Knauer, and
  Meyerhofer]{moore_observation_1995}
C.~I. Moore, J.~P. Knauer, and D.~D. Meyerhofer.
\newblock Observation of the {Transition} from {Thomson} to {Compton}
  {Scattering} in {Multiphoton} {Interactions} with {Low}-{Energy} {Electrons}.
\newblock \emph{Physical Review Letters}, 74\penalty0 (13):\penalty0
  2439--2442, March 1995.
\newblock ISSN 0031-9007, 1079-7114.
\newblock \doi{10.1103/PhysRevLett.74.2439}.

\bibitem[Salamin and Faisal(1997)]{salamin_ponderomotive_1997}
Yousef~I. Salamin and Farhad H.~M. Faisal.
\newblock Ponderomotive scattering of electrons in intense laser fields.
\newblock \emph{Physical Review A}, 55\penalty0 (5):\penalty0 3678--3683, May
  1997.
\newblock ISSN 1050-2947, 1094-1622.
\newblock \doi{10.1103/PhysRevA.55.3678}.

\bibitem[Erikson and Singh(1994)]{erikson_polarization_1994}
W.~L. Erikson and Surendra Singh.
\newblock Polarization properties of {Maxwell}-{Gaussian} laser beams.
\newblock \emph{Physical Review E}, 49\penalty0 (6):\penalty0 5778--5786, June
  1994.
\newblock ISSN 1063-651X, 1095-3787.
\newblock \doi{10.1103/PhysRevE.49.5778}.

\bibitem[Pang et~al.(2002)Pang, Ho, Yuan, Cao, Kong, Wang, Shao, Esarey, and
  Sessler]{pang_subluminous_2002}
J.~Pang, Y.~K. Ho, X.~Q. Yuan, N.~Cao, Q.~Kong, P.~X. Wang, L.~Shao, E.~H.
  Esarey, and A.~M. Sessler.
\newblock Subluminous phase velocity of a focused laser beam and vacuum laser
  acceleration.
\newblock \emph{Physical Review E}, 66\penalty0 (6):\penalty0 066501, December
  2002.
\newblock \doi{10.1103/PhysRevE.66.066501}.
\newblock Publisher: American Physical Society.

\bibitem[Longman and Fedosejevs(2022)]{longman_modeling_2022}
A.~Longman and R.~Fedosejevs.
\newblock Modeling of high intensity orbital angular momentum beams for
  laser–plasma interactions.
\newblock \emph{Physics of Plasmas}, 29\penalty0 (6):\penalty0 063109, June
  2022.
\newblock ISSN 1070-664X.
\newblock \doi{10.1063/5.0093067}.
\newblock Publisher: American Institute of Physics.

\bibitem[Wilks et~al.(1992)Wilks, Kruer, Tabak, and
  Langdon]{wilks_absorption_1992}
S.~C. Wilks, W.~L. Kruer, M.~Tabak, and A.~B. Langdon.
\newblock Absorption of ultra-intense laser pulses.
\newblock \emph{Physical Review Letters}, 69\penalty0 (9):\penalty0 1383--1386,
  August 1992.
\newblock ISSN 0031-9007.
\newblock \doi{10.1103/PhysRevLett.69.1383}.

\bibitem[Cai et~al.(2006)Cai, Yu, Zhu, and Zheng]{cai_short-pulse_2006}
Hong-bo Cai, Wei Yu, Shao-ping Zhu, and Chun-yang Zheng.
\newblock Short-pulse laser absorption via {J}×{B} heating in ultrahigh
  intensity laser plasma interaction.
\newblock \emph{Physics of Plasmas}, 13\penalty0 (11):\penalty0 113105,
  November 2006.
\newblock ISSN 1070-664X.
\newblock \doi{10.1063/1.2372463}.
\newblock Publisher: American Institute of Physics.

\bibitem[Cui et~al.(2013)Cui, Wang, Sheng, Li, and Zhang]{cui_laser_2013}
Yun-Qian Cui, Wei-Min Wang, Zheng-Ming Sheng, Yu-Tong Li, and Jie Zhang.
\newblock Laser absorption and hot electron temperature scalings in
  laser–plasma interactions.
\newblock \emph{Plasma Physics and Controlled Fusion}, 55\penalty0
  (8):\penalty0 085008, August 2013.
\newblock ISSN 0741-3335, 1361-6587.
\newblock \doi{10.1088/0741-3335/55/8/085008}.

\bibitem[noa({\natexlab{d}})]{noauthor_clpu_nodate}
{CLPU} - {Technical} {Features}, {\natexlab{d}}.
\newblock URL \url{https://www.clpu.es/en/facilties-vega-features}.

\bibitem[Chen(1984)]{chen_introduction_1984}
Francis~F. Chen.
\newblock \emph{Introduction to {Plasma} {Physics} and {Controlled} {Fusion}}.
\newblock Springer US, Boston, MA, 1984.
\newblock ISBN 978-1-4419-3201-3 978-1-4757-5595-4.
\newblock \doi{10.1007/978-1-4757-5595-4}.

\bibitem[Bellan(2008)]{bellan_fundamentals_2008}
Paul~M. Bellan.
\newblock \emph{Fundamentals of {Plasma} {Physics}}.
\newblock Cambridge University Press, July 2008.
\newblock ISBN 978-1-139-44973-1.
\newblock Google-Books-ID: v2dER3SUrtsC.

\bibitem[Takahaahi et~al.()Takahaahi, Kohda, and
  Niyaltara]{takahaahi_mechanism_nodate}
Kenji Takahaahi, Katauhiro Kohda, and Junji Niyaltara.
\newblock {MECHANISM} {OF} {PHOTOSTIMULATEO} {LUMINESCENCE} {IN} {BaFX}:{Eu}
  ({X}{\textasciitilde}{Cl},{Nr}) {PHOSPHORS}.
\newblock page~3.

\bibitem[Takahashi et~al.(1985)Takahashi, Miyahara, and
  Shibahara]{takahashi_photostimulated_1985}
K.~Takahashi, J.~Miyahara, and Y.~Shibahara.
\newblock Photostimulated {Luminescence} ({PSL}) and {Color} {Centers} in
  {BaFX} : {Eu2} + ( {X} = {Cl} , {Br} , {I} ) {Phosphors}.
\newblock \emph{Journal of The Electrochemical Society}, 132\penalty0
  (6):\penalty0 1492, June 1985.
\newblock ISSN 1945-7111.
\newblock \doi{10.1149/1.2114149}.
\newblock Publisher: IOP Publishing.

\bibitem[Miyahara et~al.(1986)Miyahara, Takahashi, Amemiya, Kamiya, and
  Satow]{miyahara_new_1986}
Junji Miyahara, Kenji Takahashi, Yoshiyuki Amemiya, Nobuo Kamiya, and Yoshinori
  Satow.
\newblock A new type of {X}-ray area detector utilizing laser stimulated
  luminescence.
\newblock \emph{Nuclear Instruments and Methods in Physics Research Section A:
  Accelerators, Spectrometers, Detectors and Associated Equipment},
  246\penalty0 (1):\penalty0 572--578, May 1986.
\newblock ISSN 0168-9002.
\newblock \doi{10.1016/0168-9002(86)90156-7}.

\bibitem[noa({\natexlab{e}})]{noauthor_stopping_nodate}
Stopping {Power} and {Range} {Tables} for {Electrons}, {\natexlab{e}}.
\newblock URL \url{https://physics.nist.gov/cgi-bin/Star/e_table.pl}.

\bibitem[Boutoux et~al.(2015)Boutoux, Rabhi, Batani, Binet, Ducret, Jakubowska,
  Nègre, Reverdin, and Thfoin]{boutoux_study_2015}
G.~Boutoux, N.~Rabhi, D.~Batani, A.~Binet, J.-E. Ducret, K.~Jakubowska, J.-P.
  Nègre, C.~Reverdin, and I.~Thfoin.
\newblock Study of imaging plate detector sensitivity to 5-18 {MeV} electrons.
\newblock \emph{Review of Scientific Instruments}, 86\penalty0 (11):\penalty0
  113304, November 2015.
\newblock ISSN 0034-6748.
\newblock \doi{10.1063/1.4936141}.

\bibitem[Williams et~al.(2014)Williams, Maddox, Chen, Kojima, and
  Millecchia]{williams_calibration_2014}
G.~Jackson Williams, Brian~R. Maddox, Hui Chen, Sadaoki Kojima, and Matthew
  Millecchia.
\newblock Calibration and equivalency analysis of image plate scanners.
\newblock \emph{Review of Scientific Instruments}, 85\penalty0 (11):\penalty0
  11E604, November 2014.
\newblock ISSN 0034-6748.
\newblock \doi{10.1063/1.4886390}.
\newblock Publisher: American Institute of Physics.

\bibitem[Bonnet et~al.(2013)Bonnet, Comet, Denis-Petit, Gobet, Hannachi,
  Tarisien, Versteegen, and Aléonard]{bonnet_response_2013}
T.~Bonnet, M.~Comet, D.~Denis-Petit, F.~Gobet, F.~Hannachi, M.~Tarisien,
  M.~Versteegen, and M.~M. Aléonard.
\newblock Response functions of imaging plates to photons, electrons and {4He}
  particles.
\newblock \emph{Review of Scientific Instruments}, 84\penalty0 (10):\penalty0
  103510, October 2013.
\newblock ISSN 0034-6748.
\newblock \doi{10.1063/1.4826084}.

\bibitem[noa({\natexlab{f}})]{noauthor_mppcs_nodate}
{MPPCs} ({SiPMs}) / {MPPC} arrays {\textbar} {Hamamatsu} {Photonics},
  {\natexlab{f}}.
\newblock URL
  \url{https://www.hamamatsu.com/us/en/product/optical-sensors/mppc/mppc_mppc-array.html}.

\bibitem[Wilhelm and Durfee(2019)]{wilhelm_tilted_2019}
Alex~M. Wilhelm and Charles~G. Durfee.
\newblock Tilted {Snowplow} {Ponderomotive} {Electron} {Acceleration} {With}
  {Spatio}-{Temporally} {Shaped} {Ultrafast} {Laser} {Pulses}.
\newblock \emph{Frontiers in Physics}, 7, 2019.
\newblock ISSN 2296-424X.
\newblock \doi{10.3389/fphy.2019.00066}.

\end{thebibliography}

\end{document}